\documentclass[authoryear]{elsarticle}

\usepackage{geometry}
\usepackage{bm}
\usepackage{graphicx}
\usepackage{enumitem}
\usepackage{etoolbox}
\usepackage{amssymb}
\usepackage[linesnumbered,ruled,vlined]{algorithm2e}
\usepackage{multirow}
\usepackage{placeins}
\usepackage{amsmath}
\usepackage{mathtools}
\usepackage{makecell}
\usepackage{threeparttable}
\usepackage{amsthm}
\usepackage[colorlinks]{hyperref}
\usepackage[title]{appendix}
\usepackage{booktabs}
\usepackage{float}
\usepackage{subcaption}
\usepackage{comment}

\newcommand{\tablebodyfont}{\small}
\AtBeginEnvironment{tabular}{\tablebodyfont}
\theoremstyle{definition}

\theoremstyle{remark}

\begin{document}

\begin{frontmatter}
\title{SelectLight: Learning to Select Signal Plans Generated by Distributed Model Predictive Control for Urban Traffic Networks}
\author[1]{Lyuzhou Luo}
\ead{jexxllz@tongji.edu.cn}
\author[2]{Chaopeng Tan\corref{cor1}}
\ead{chaopeng.tan@tu-dresden.de}
\author[1]{Zhengyong Gao}
\ead{gao00@tongji.edu.cn}
\author[1]{Hong Zhu}
\ead{hongzhu1990@tongji.edu.cn}
\author[3]{Andrea D'Ariano}
\ead{andrea.dariano@uniroma3.it}
\author[1]{Keshuang Tang\corref{cor1}}
\ead{tang@tongji.edu.cn}
\cortext[cor1]{Corresponding authors.}
\address[1]{Key Laboratory of Road and Traffic Engineering of the Ministry of Education, College of Transportation, Tongji University, Cao'an Road 4800, Shanghai 201804, China}
\address[2]{Chair of Traffic Process Automation, Technische Universit\"at Dresden, Hettnerstra{\ss}e 3, 01069, Dresden, Germany}
\address[3]{Department of Civil, Computer Science and Aeronautical Technologies Engineering, Roma Tre University, Via della Vasca Navale 79, 00146 Rome, Italy}

\begin{abstract}
Coordinated traffic signal control across urban networks must adapt to changing demand while satisfying operational constraints. Multi-objective distributed model predictive control (DMPC) can construct feasible signal plans online, but prescribed rules for selecting among trade-off solutions cannot learn from realized closed-loop outcomes. We propose SelectLight, which implements post-optimization selection by allowing a multi-agent reinforcement learning (MARL) policy to choose directly from plans generated online by DMPC. At each control update, state-pruned multi-objective dynamic programming (SP-MODP) evaluates plans with a Newellian point--spatial queue model and returns a bounded set of mutually nondominated candidate signal plans for total queueing delay, peak queue accumulation, and total number of stops. A topology-aware attention policy trained with independent proximal policy optimization (IPPO) selects one unmodified plan from each variable-size set. This confines learning to candidate selection, preserves the prescribed signal timing constraints, and leaves the selected plan and its predicted objective trade-offs available for inspection. Experiments on two 28-intersection SUMO networks show that SelectLight achieves the best delay-related performance and that its advantage widens with demand. At twice the baseline demand, it reduces queueing delay and waiting time by 5.57\% and 6.44\%, respectively, relative to the strongest baseline. SelectLight also incurs the lowest transfer loss under every tested demand shift. With a 120~s prediction horizon, the per-intersection 99th-percentile SP-MODP solution time is 5.408~ms, well below the 5~s control interval.
\end{abstract}

\end{frontmatter}

\section{Introduction}
\label{sec:introduction}

Network-level urban traffic signal control affects congestion and travel delay by allocating right-of-way among competing traffic movements \citep{papageorgiou2003review,qadri2020stateofart,wang2023critical,li2023survey,tan2025connected}. At an isolated intersection, this allocation determines how local queues are served; across a network, it also shapes downstream arrivals, queue propagation, and platoon progression \citep{tan2026leveraging}. A network controller must therefore coordinate neighboring interactions, satisfy signal timing constraints, and respond to changing demand. Different control paradigms meet these requirements through different mechanisms. Actuated and max pressure (MP) controllers react to measured demand or traffic state imbalance without requiring an explicit prediction horizon \citep{qadri2020stateofart,varaiya2013max,tan2026cvmp,tan2026privatemp}, whereas predictive controllers use prediction of future arrivals to evaluate signal decisions \citep{zhu2025asynchronous}.

Model predictive control (MPC) makes finite-horizon prediction and signal timing constraints explicit. Distributed model predictive control (DMPC) decomposes network control into coupled local MPC problems, reducing the need for a single network-wide optimization \citep{aboudolas2010rollinghorizon,ye2014distributed,pham2023distributed,xin2023model,wu2024distributed}. Because MPC repeatedly optimizes from the current estimated state, it adapts its plan online \citep{rinaldi2019mixed} and can generalize beyond previously encountered states when its prediction model remains valid \citep{reiter2026synthesis}. At the same time, traffic signal control typically requires trade-offs among multiple objectives. For example, traffic management priorities may shift from queue length reduction during peak hours to minimizing total delay during off-peak periods, even though the system-level objective is to minimize total network travel delay \citep{luo2025distributed}.

Multi-objective MPC formalizes this trade-off directly within the optimization. Weighted formulations specify their relative importance before optimization and return one plan \citep{alislam2017distributed}; Pareto-based formulations instead retain plans with different predicted trade-offs and apply a prescribed rule to select one for execution \citep{li2019multiobjective,luo2025distributed}. In the latter case, the candidate plans and their objective values change with the traffic state, but the selection rule does not learn from realized outcomes. Moreover, MPC executes only the initial part of a selected plan. Prediction errors, interactions with neighboring intersections, and subsequent replanning can therefore make its realized closed-loop performance differ from its finite-horizon evaluation.

Reinforcement learning (RL) provides a complementary mechanism by learning decision rules from repeated interaction with traffic. Multi-agent reinforcement learning (MARL) extends this formulation to networks through decentralized policies and learned coordination \citep{wang2023critical,chu2020multiagent,wei2019colight}. Recent RL controllers augment the policy with predicted traffic states \citep{han2023mitigating,ye2025deep,tang2024traffic}, multiple operational criteria \citep{fang2023multiobjective,saiki2023flexible}, or high-level selection among objective-specialized sub-policies \citep{xu2021hierarchically}. Across these designs, the deployed policy ultimately chooses a phase, a green-duration vector, or a learned sub-policy. This action structure supports low-cost policy inference and allows decisions to reflect realized returns, but it does not perform online constrained construction and evaluation of finite-horizon signal plans. It also gives traffic managers limited visibility into why one control is preferred over another, complicating operational scrutiny of learned decisions. In addition, performance under unseen conditions remains dependent on training coverage, and fixed policies may require adaptation when the operating distribution changes \citep{wang2023critical,reiter2026synthesis}.

The complementary properties of MPC and RL motivate deployment-time hybrids in which predictive optimization remains active while a learned component influences the applied control \citep{reiter2026synthesis}. Existing designs either configure the current MPC problem before it is solved or correct one control decision after optimization \citep{sun2024adaptive,zarrouki2024safe,sun2024novel,remmerswaal2022combined}. These interfaces expose an optimization specification or one MPC output rather than the current set of MPC-generated signal plans. We formulate \emph{post-optimization selection}, in which MPC constructs a state-dependent set of candidate plans and a learned policy selects one without modifying it. Because the number and content of available plans vary at every control update, the policy must compare the current candidate plan features rather than assign fixed semantics to candidate indices. This interface confines learning to candidate selection while leaving the selected plan and its predicted objective trade-offs available for inspection. Figure~\ref{fig:mpcrl-integration-patterns} summarizes the deployment-time integration patterns considered in this paper.

We instantiate post-optimization selection in \emph{SelectLight}, which couples DMPC candidate generation with an attention-based selector trained using independent proximal policy optimization (IPPO) \citep{schulman2017proximal,dewitt2020independent}. At each control update, a prediction module and per-intersection candidate generators jointly implement coupled local multi-objective MPC problems. State-pruned multi-objective dynamic programming (SP-MODP) solves each local problem using a Newellian point--spatial queue model \citep{newell2002simplified,wang2024traffic} and returns a bounded set of mutually nondominated candidate signal plans characterized by three predicted objectives: total queueing delay, peak queue accumulation, and total number of stops. Multi-objective optimization is used here to construct informative candidates; it is not a prerequisite of the post-optimization selection pattern. A topology-aware policy encodes the local traffic state and each candidate plan, then uses masked scaled dot-product attention to compare only the available candidates. The policy is trained from realized local queueing delay over successive executed intervals. Because every candidate plan satisfies the prescribed signal timing constraints and the policy does not modify its choice, the selected plan satisfies them as well.

This paper makes the following contributions:
\begin{itemize}
  \item We propose post-optimization selection as a new deployment-time MPC--RL integration pattern for network traffic signal control, in which local MPC problems construct state-dependent sets of candidate signal plans and a learned candidate selection policy selects one plan for execution without modifying it. The pattern retains control-time predictive planning while restricting the policy to candidate plans that satisfy the prescribed signal timing constraints. The policy learns the selection rule from realized closed-loop outcomes, while the selected plan and its predicted objective trade-offs remain available for inspection.

  \item We implement this pattern using SP-MODP for candidate generation and a topology-aware attention policy for candidate selection. Relative to the MODP in \citet{luo2025distributed}, SP-MODP adapts the label state representation to the Newellian point--spatial queue model. Objective space pruning and a per-node label cap bound the partial plans propagated through the search. Rather than applying a prescribed rule to select one terminal plan, SP-MODP returns multiple plans for learned selection. Shared candidate encoding and masked attention support variable-size candidate sets without assigning persistent meaning to candidate indices.

  \item We demonstrate that SelectLight transfers its learned candidate-selection rule across unseen demand levels and network configurations, while online MPC regenerates the candidate signal plans for each deployment condition. Topology-aware encoding and parameter sharing allow the selector to operate across heterogeneous intersections. Experiments on two 28-intersection urban networks show increasing delay reductions under heavier demand and lower zero-shot transfer loss than direct RL policies, without fine-tuning.
\end{itemize}

The remainder of the paper is organized as follows. Section~\ref{sec:related-work} reviews MPC-based and RL-based signal control, identifies the research gap, and positions post-optimization selection within deployment-time MPC--RL integration. Section~\ref{sec:problem-formulation} formulates the lower-level candidate generation and upper-level selection problems. Section~\ref{sec:methodology} presents the SelectLight framework. Sections~\ref{sec:traffic-prediction}, \ref{sec:lower-level-dmpc-solver}, and~\ref{sec:upper-level-rl-policy} detail arrival prediction and traffic state estimation, multi-objective MPC candidate generation, and attention-based candidate selection, respectively. Section~\ref{sec:experiments} reports the experiments, and Section~\ref{sec:conclusion} concludes the paper.

\section{Related work}
\label{sec:related-work}

\subsection{MPC-based network traffic signal control}

MPC optimizes signal decisions over a finite prediction horizon subject to predicted traffic dynamics and operational constraints. Early network-level studies developed store-and-forward formulations for large congested networks and receding horizon quadratic programs for signal timing \citep{aboudolas2009storeandforward,aboudolas2010rollinghorizon}. Later formulations extended this paradigm to integrated urban--freeway control, time-varying mixed-integer signal models, and adaptive linear network models \citep{berg2007integrated,guilliard2016nonhomogeneous,wang2022optimizing}. These centralized formulations can represent network interactions within one optimization problem, but their computational burden increases with the number of coupled traffic states, signal variables, and operational constraints.

DMPC addresses this scalability issue by decomposing network control into local MPC problems while coordinating them through exchanged traffic or control information. Representative formulations coordinate local signal-split problems through dual decomposition \citep{ye2014distributed}, exchange predicted sending flows and receiving capacities between neighboring controllers \citep{alislam2017distributed}, or use the alternating direction method of multipliers to handle stochastic network flows \citep{pham2023distributed}. Repeated state-conditioned optimization also gives MPC online adaptability when the prediction model remains valid \citep{reiter2026synthesis}. DMPC therefore retains finite-horizon prediction and explicit signal optimization without requiring a single network-wide solve, although its coordination depends on the information supplied to and exchanged among the local problems.

Traffic signal control often involves objectives that characterize different operational effects, making the representation and resolution of their trade-offs an important design choice. Weighted formulations specify the relative importance of these effects before optimization and return one solution \citep{alislam2017distributed}. Pareto-based formulations instead retain multiple trade-off solutions during the search. A real-time intersection controller uses particle swarm optimization to approximate the Pareto front \citep{jiao2016pareto}, while a spatial MPC formulation combines multi-objective particle swarm optimization with a preference-ranking rule for oversaturated networks \citep{ma2021backpressurebased}. The predictive controller in \citet{li2019multiobjective} uses a genetic algorithm to generate online Pareto solutions and selects knee or extreme solutions according to the operating case. At the distributed network scale, the formulation in \citet{luo2025distributed} generates nondominated local signal plans through multi-objective dynamic programming and applies a rule-based local selector. These studies establish online generation of multiple predictive signal plans, but their executable plan remains determined by a prescribed preference rule rather than a rule learned from realized closed-loop outcomes.

\subsection{RL-based traffic signal control}

RL replaces repeated online solution of an explicit signal-planning problem with a policy learned through interaction with traffic. Early applications learned phase or duration decisions at individual intersections using actor--critic or value-based methods \citep{aslani2017adaptive,wei2018intellilight}. Because traffic propagation couples neighboring intersections, later work formulated network control as MARL. MA2C incorporates local and neighboring observations, MPLight combines pressure-based rewards with parameter sharing, and CoLight uses graph attention to learn the relative influence of neighboring intersections \citep{chu2020multiagent,chen2020thousand,wei2019colight}. These methods provide representative mechanisms for decentralized network-level control and coordination.

Recent RL controllers enrich the information and coordination structure available to the learned policy. Predictive approaches provide a policy with the traffic state expected after a signal transition interval or with historical, current, and predicted traffic flow features \citep{han2023mitigating,ye2025deep}. A related receding horizon framework combines a graph-based traffic predictor with a deep Q-network that directly selects predefined phases \citep{tang2024traffic}. MetaLight targets rapid adaptation across traffic scenarios, model-based graph RL uses learned dynamics to support inductive transfer, and HALO introduces hierarchical guidance for large-scale control \citep{zang2020metalight,devailly2024modelbased,zhu2026halo}. Related regional control studies apply RL to adaptive perimeter control and multi-region metering \citep{chen2022data,zhou2023scalable}. Taken together, these studies enrich RL signal control with predicted traffic information, cross-scenario adaptation and transfer, hierarchical guidance, and coordination at intersection and regional scales.

Multiple objectives have also been incorporated into RL through reward design, conditioned policies, and hierarchical policy selection. NACRL combines efficiency, safety, and coordination terms in a comprehensive reward for decentralized phase control \citep{fang2023multiobjective}. A weight-conditioned multi-objective RL formulation learns a convex coverage set of policies and assigns objective weights online from the observed directional flow ratio \citep{saiki2023flexible}. HiLight instead uses a high-level controller to choose among sub-policies specialized for queue length, waiting time, and delay \citep{xu2021hierarchically}. In each case, the controller ultimately performs phase-level actions through one learned policy or a selected learned sub-policy.

Overall, RL-based signal controllers learn decision rules from closed-loop experience and execute them through policy inference. Predicted traffic features, learned dynamics, multiple reward components, and policy hierarchies enrich these controllers, but their deployed policies still choose direct signal actions or learned sub-policies rather than controls constructed through online constrained optimization. Performance outside the training distribution also remains dependent on the coverage and fidelity of the training environment \citep{wang2023critical,reiter2026synthesis}.

\subsection{Deployment-time MPC--RL integration}

MPC--RL combinations differ in the role assigned to MPC and in whether MPC remains active after learning. The taxonomy of \citet{reiter2026synthesis} distinguishes MPC as an expert actor, a component of the deployed policy, or part of the critic. When MPC serves only as an expert actor or critic, it is used during learning and removed before deployment. For example, critic-based methods use MPC-derived action-value functions to train standalone policies \citep{carius2020mpcnet,ghezzi2023imitation}. In traffic signal control, \citet{guo2021integrated} use MPC-generated trajectories to initialize policy and value networks before PPO training, then deploy the trained policy without MPC at execution time. In these examples, MPC influences policy learning but no longer participates in control after training.

This section focuses on deployment-time integration patterns that satisfy two conditions: MPC remains active during controller execution after training and constructs or evaluates a state-conditioned control at each update, and an RL-trained component influences the applied control. If MPC is removed after learning, the two mechanisms can no longer jointly determine the control at execution time. The central question is therefore where the learned decision enters relative to the current MPC solve and what decision object it exposes at the MPC--RL interface. We compare the reviewed designs along these two dimensions, independent of their control objectives and optimization algorithms.

One pattern identified in the reviewed deployment-time designs is \emph{pre-optimization configuration}, in which RL intervenes before MPC solves the current optimization problem. Outside transportation operations, RL has been used to tune economic MPC schemes and state-dependent MPC meta-parameters \citep{gros2020datadriven,bohn2023optimization}, select discrete mode sequences that restrict mixed-integer MPC problems \citep{silva2025integrating}, and regulate MPC weights and horizons for battery thermal management \citep{li2025hyperparameter}. Transportation applications adapt components of parameterized MPC for freeway traffic control \citep{sun2024adaptive}, learn a parameterized MPC policy for highway ramp metering \citep{airaldi2025reinforcement}, or adapt stochastic-MPC robustness and look-ahead settings for autonomous-vehicle motion control \citep{zarrouki2024adaptivestochastic}. The structurally closest example uses PPO to select an MPC weight configuration from an offline Pareto catalog in that setting \citep{zarrouki2024safe}. Across these applications, the MPC--RL decision interface carries an optimization specification or restriction from RL to MPC, and its effect on the applied control is mediated by the subsequent MPC solve. MPC constructs the applied control within the configured problem. This pattern is well suited to adapting priorities, model behavior, or computational structure while preserving an optimizer-centered decision process.

A second reviewed pattern is \emph{post-optimization correction}, in which MPC first returns a baseline control decision and RL then modifies it. In energy-management applications, learned compensation has been applied to DMPC or MPC baseline actions for shipboard power systems and fuel-cell/battery hybrid energy systems \citep{fu2024compensation,liu2025energymanagement}. Transportation applications include additive corrections to MPC baseline actions for freeway traffic management \citep{sun2024novel}, a bounded residual policy over an MPC backbone for autonomous-vehicle platooning \citep{allahloh2026residual}, and learned compensatory inputs to hierarchical MPC for eco-driving \citep{liu2026hierarchical}. The closest urban traffic signal control study combines MPC-computed green-time percentages with discrete corrections selected by RL in a two-intersection network \citep{remmerswaal2022combined}. Here, the MPC--RL decision interface carries one baseline control from MPC to RL, and the learned action is a residual or transformation applied to that control. The final control is therefore jointly determined by the baseline and the learned correction. This pattern is well suited to compensating for model mismatch or systematic errors in the MPC baseline while retaining it as an operating reference.

\subsection{Research gap and positioning}
\label{sec:research-gap-positioning}

The preceding studies reveal a mismatch between the candidate signal plans constructed by online optimization and the choices available to a learned policy. Pareto-based MPC can retain multiple feasible trade-off plans, but a prescribed rule determines which plan is applied \citep{ma2021backpressurebased,li2019multiobjective,luo2025distributed}. RL signal controllers instead choose direct signal actions or learned sub-policies \citep{tang2024traffic,xu2021hierarchically}. Among deployment-time hybrids, learning supplies an optimization specification under pre-optimization configuration and modifies one optimized output under post-optimization correction \citep{zarrouki2024safe,remmerswaal2022combined}. The state-dependent candidate set generated by the current MPC solve therefore remains outside the learned decision loop.

\begin{figure}
\centering
\includegraphics[width=0.90\linewidth]{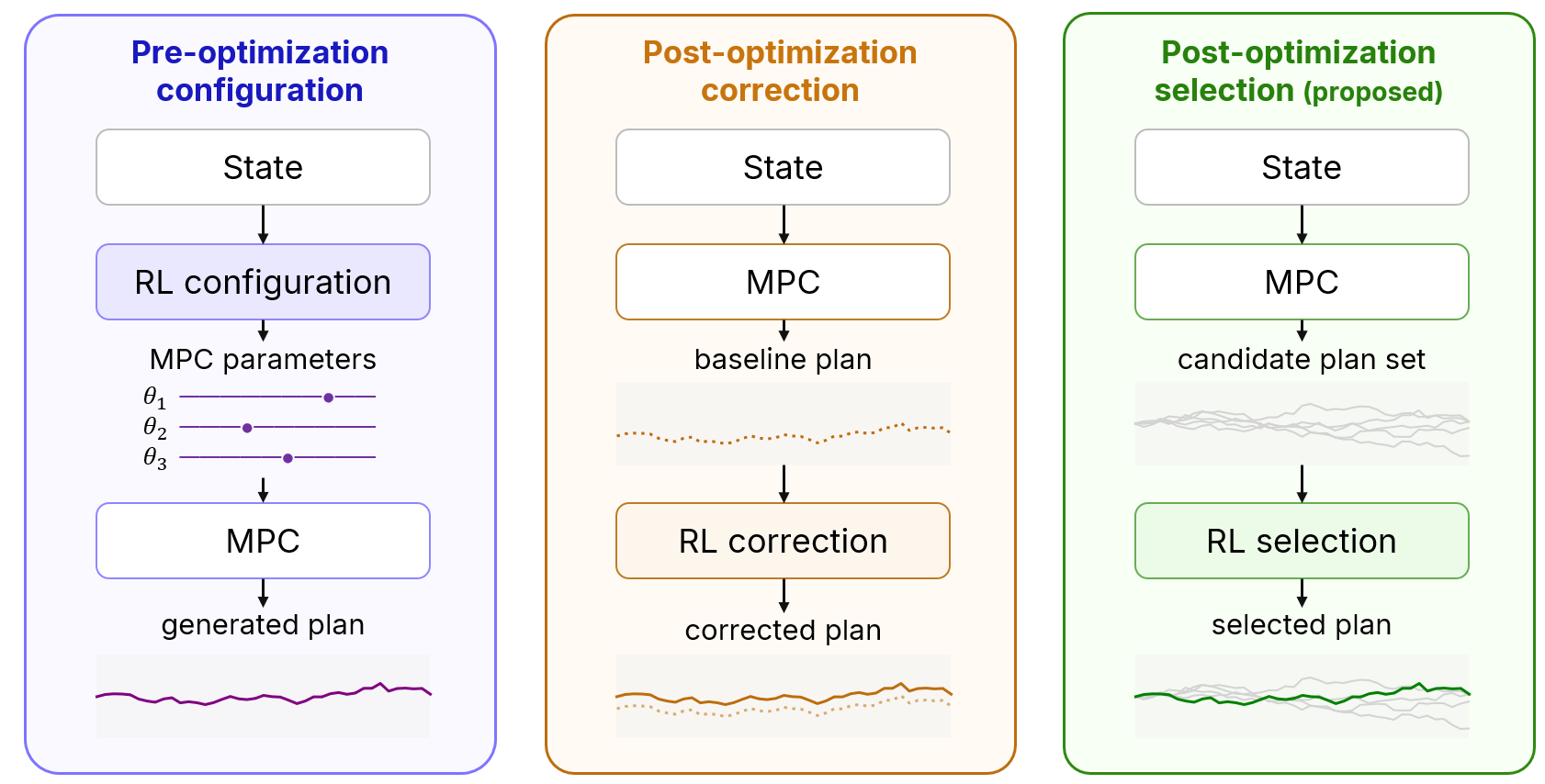}
\caption{Three deployment-time MPC--RL integration patterns distinguished by the decision object exposed at their interface. RL configures the problem subsequently solved by MPC, transforms one MPC output, or, in the proposed pattern, selects an unmodified member of a state-dependent control set constructed by MPC.}
\label{fig:mpcrl-integration-patterns}
\end{figure}

We formulate this missing interface as \emph{post-optimization selection}. At each control update, local multi-objective MPC problems generate bounded, variable-size sets of feasible signal plans, and an upper-level policy selects one plan without modifying it. Because the available plans change with the traffic state, the policy compares their features rather than assigning persistent meaning to candidate indices. This separation restricts the policy to signal plans that satisfy the prescribed signal timing constraints while allowing the selection rule to learn from closed-loop outcomes. Figure~\ref{fig:mpcrl-integration-patterns} summarizes this pattern alongside pre-optimization configuration and post-optimization correction.

\section{Problem formulation}
\label{sec:problem-formulation}

Consider an urban traffic network represented by a directed graph $\mathcal{G}=(\mathcal{V},\mathcal{E})$, where $\mathcal{V}$ is the set of signalized intersections and $\mathcal{E}$ is the set of directed links. Each intersection $i\in\mathcal{V}$ is controlled by an agent. Let $\mathcal{N}_i$ denote the set of adjacent intersections that can affect traffic arriving at intersection $i$, and let $\mathcal{M}_i$ denote the set of inbound lanes controlled by agent $i$.

The system-level objective is to minimize total network travel delay subject to traffic dynamics and signal control constraints. To support online decision-making, we formulate a hierarchical control problem with two decision levels. Because complete trip delay is not available within each short decision interval, the two levels use congestion-related online surrogates. The lower level evaluates three predicted objectives---total queueing delay, peak queue accumulation, and total number of stops---whereas the upper level uses negative realized local queueing delay as its reward. These signals guide online decisions, while total network travel delay remains the primary system-level objective and evaluation criterion.

At the lower level, control is formulated using DMPC, with one local multi-objective finite-horizon MPC problem assigned to each intersection. Each local problem produces a compact set of feasible signal plans, termed candidate signal plans, that represent different objective trade-offs. Predictive arrival profiles couple these local problems across adjacent intersections. At the upper level, the candidate selection policy chooses one candidate plan based on the current and predicted local traffic conditions. This section formulates the two decision problems; their traffic models, solution procedure, and policy implementation are introduced in the subsequent sections.

\subsection{Temporal structure}

The framework operates at two timescales. Let $\Delta_k$ be the micro-step duration. The signal decision interval contains $M$ micro steps, so macro step $t$ begins at micro step $k_t=tM$. The lower level optimizes over a horizon of $H$ micro steps, where $H>M$.

At each macro step, the following procedure is performed:
\begin{enumerate}
    \item \textbf{State update:} Each agent obtains its current local state $x_i(k_t)$ and predictive arrival profile $\hat{\mathbf q}_i^{\mathrm{arr}}(p\mid k_t)$.
    \item \textbf{Candidate generation:} Each agent solves its local multi-objective MPC problem to generate a bounded set of candidate signal plans over the prediction horizon.
    \item \textbf{Candidate selection:} The upper-level policy constructs a local observation and selects one candidate plan.
    \item \textbf{Receding horizon execution:} Only the first $M$ micro steps of the selected plan are applied. Its unexecuted tail informs the prediction at the next macro step and is replaced after replanning.
\end{enumerate}

\begin{figure}
    \centering
    \includegraphics[width=\linewidth]{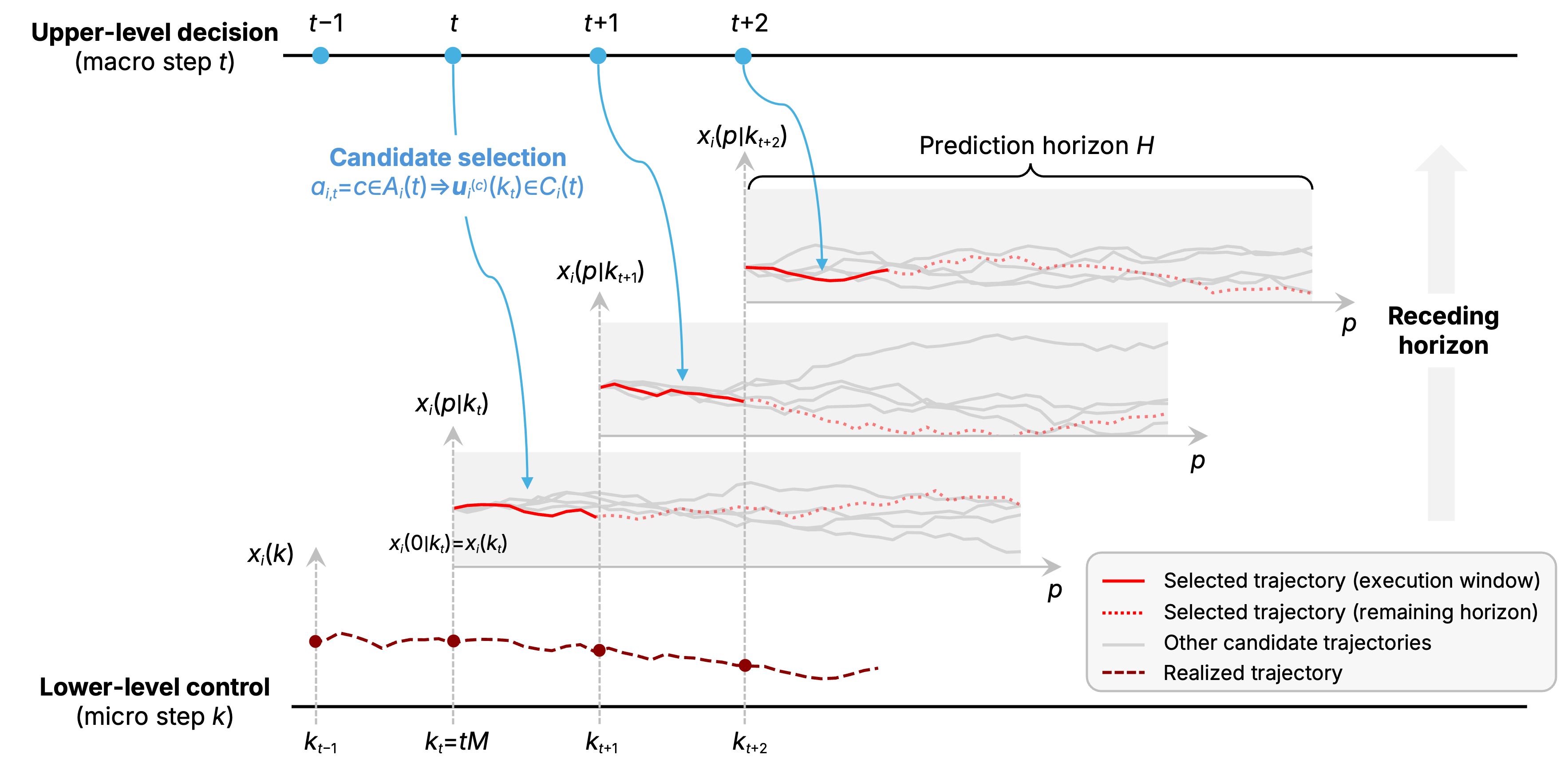}
    \caption{Temporal structure of the hierarchical control framework. At each macro step, each agent solves its local multi-objective MPC problem to generate candidate signal plans over $H$ micro steps, the upper level selects one candidate plan, and only its first $M$ micro steps are executed. The unexecuted plan tail informs the next prediction and is replaced when the horizon recedes.}
    \label{fig:temporal-structure}
\end{figure}

This procedure is repeated in a receding horizon manner as illustrated in Figure~\ref{fig:temporal-structure}.

\subsection{Lower-level multi-objective DMPC}
\label{sec:lower-level-dmpc-formulation}

The lower-level DMPC formulation consists of coupled local multi-objective MPC problems, one for each intersection. For intersection $i$, let $x_i(k_t)$ denote the current local traffic state, $\hat{\mathbf q}_i^{\mathrm{arr}}(p\mid k_t)$ the predictive arrival profile, and $\mathbf u_i(p\mid k_t)$ the signal control vector at prediction step $p$. Their model-specific constructions are deferred to Sections~\ref{sec:traffic-prediction} and~\ref{sec:lower-level-dmpc-solver}.

The predictive arrival profile couples the local MPC problems. Let $\mathcal{Y}(k_t)$ collect the traffic observations and boundary departure forecasts available at macro step $t$, and let $\mathbf u_j^{\mathrm{sel}}(k_{t-1})$ denote the plan selected by upstream agent $j$ at the preceding macro step. The profile supplied to intersection $i$ can be written as
\begin{equation}
    \hat{\mathbf q}_i^{\mathrm{arr}}(\cdot\mid k_t)
    =
    \mathcal{H}_i^{\mathrm{arr}}
    \left(
        \mathcal{Y}(k_t),
        \left\{
            \mathbf u_j^{\mathrm{sel}}(k_{t-1})
        \right\}_{j\in\mathcal{N}_i}
    \right),
    \label{eq:dmpc-coupling}
\end{equation}
where $\mathcal{H}_i^{\mathrm{arr}}$ propagates observed and forecast departures, together with departures implied by upstream plans, to the inbound lanes of intersection $i$. After candidate selection, the newly selected plan affects the arrival profiles of downstream agents at the next macro step. Thus, the local finite-horizon problems are coupled through previously selected plans and their induced arrival profiles. Collectively, these coupled local MPC problems constitute the lower-level DMPC formulation, while each optimization remains local to one intersection. Section~\ref{sec:traffic-prediction} specifies the propagation and feedback operations represented by $\mathcal{H}_i^{\mathrm{arr}}$.

At macro step $t$, the local MPC subproblem associated with agent $i$ is formulated as follows:
\begin{equation}
\begin{aligned}
    \underset{\mathbf u_i(k_t)}{\operatorname{ParetoMin}}
    \quad&
    \mathbf F_i(k_t)
    =
    \begin{bmatrix}
        F_i^{\mathrm{delay}}(k_t) \\
        F_i^{\mathrm{queue}}(k_t) \\
        F_i^{\mathrm{stop}}(k_t)
    \end{bmatrix}
    \\
    \text{s.t.}\quad
    &
    x_i(0\mid k_t)=x_i(k_t),
    \\
    &
    x_i(p+1\mid k_t)
    =
    f_i\!\left(
        x_i(p\mid k_t),
        \mathbf u_i(p\mid k_t),
        \hat{\mathbf q}_i^{\mathrm{arr}}(p\mid k_t)
    \right),
    \quad p=0,\ldots,H-1,
    \\
    &
    x_i(p+1\mid k_t)\in\mathcal{X}_i,
    \quad p=0,\ldots,H-1,
    \\
    &
    \mathbf u_i(k_t)\in\mathcal{U}_i(k_t).
\end{aligned}
\label{eq:p1}
\end{equation}
where
\[
    \mathbf u_i(k_t)
    =
    \left\{
        \mathbf u_i(p\mid k_t)
    \right\}_{p=0}^{H-1}
\]
is a signal plan over the prediction horizon. The sets $\mathcal{X}_i$ and $\mathcal{U}_i(k_t)$ encode state feasibility and signal timing constraints, respectively. The three objectives measure total queueing delay, peak queue accumulation, and total number of stops over the horizon.

Problem~\eqref{eq:p1} is the local MPC subproblem for intersection $i$ within the lower-level DMPC formulation and defines its ideal Pareto-optimal plan set. For each intersection, a bounded approximation of this set is provided to the upper level as a mutually nondominated feasible candidate set
\begin{equation}
    \mathcal{C}_i(t)
    =
    \left\{
        \mathbf u_i^{(1)}(k_t),
        \ldots,
        \mathbf u_i^{(K'_i(t))}(k_t)
    \right\},
    \qquad
    K'_i(t)\leq K,
    \label{eq:candidate-set}
\end{equation}
where $K$ is the prescribed maximum candidate set size. The concrete traffic model and candidate generation procedure are presented in Section~\ref{sec:lower-level-dmpc-solver}.

\subsection{Upper-level candidate selection problem}

The upper level resolves the combinatorial signal planning problem by selecting one candidate plan at each intersection. It is formulated as a partially observable Markov game (POMG)
\begin{equation}
    \mathcal{G}_{\mathrm{POMG}}
    =
    \left\langle
        \mathcal{I},
        \mathcal{S},
        \{\mathcal{O}_i\}_{i\in\mathcal{I}},
        \{\mathcal{A}_i\}_{i\in\mathcal{I}},
        \mathcal{P},
        \mathcal{Z},
        \{r_i\}_{i\in\mathcal{I}},
        \gamma
    \right\rangle,
\end{equation}
with the following components:
\begin{itemize}
    \item $\mathcal{I}=\mathcal{V}$ is the set of signal control agents.
    \item $\mathcal{S}$ is the global traffic state, including network-wide queue states and signal states.
    \item $\mathcal{O}_i$ is the local observation space of agent $i$, summarizing its current and predicted traffic conditions together with information about the currently available candidate plans.
    \item $\mathcal{A}_i=\{1,\ldots,K\}$ is the candidate index set under the prescribed candidate set size bound. At macro step $t$, the available action set is
    \begin{equation}
        \mathcal{A}_i(t)
        =
        \{1,\ldots,K'_i(t)\}.
    \end{equation}
    Selecting $a_{i,t}=c$ commits to $\mathbf u_i^{\mathrm{sel}}(k_t)=\mathbf u_i^{(c)}(k_t)$. The first $M$ micro steps are executed, while predicted departures associated with the unexecuted plan tail contribute to the next arrival profile update.
    \item $\mathcal{P}(s_{t+1}\mid s_t,\mathbf a_t)$ is the transition kernel induced by traffic dynamics and the jointly selected plans.
    \item $\mathcal{Z}:\mathcal{S}\rightarrow\prod_{i\in\mathcal{I}}\mathcal{O}_i$ is the observation function.
    \item $r_i$ is the local reward function. Its realization $r_{i,t}$ at macro step $t$ is the negative realized local queueing delay over the executed interval and provides an immediately measurable online surrogate for the system-level travel-delay objective.
    \item $\gamma\in[0,1)$ is the discount factor.
\end{itemize}

The implemented parameter-sharing policy, detailed in Section~\ref{sec:upper-level-rl-policy}, is denoted by $\pi_{\theta_\pi}(a_{i,t}\mid o_{i,t})$. With the team reward defined as $R_t=\sum_{i\in\mathcal I}r_{i,t}$, the cooperative objective is
\begin{equation}
    \max_{\theta_\pi}
    \;
    \mathbb{E}
    \left[
        \sum_{t=0}^{\infty}
        \gamma^t
        R_t
    \right].
    \label{eq:upper-objective}
\end{equation}
The local observation construction, policy parameterization, reward implementation, and learning algorithm are detailed in Section~\ref{sec:upper-level-rl-policy}.

\section{SelectLight framework}
\label{sec:methodology}
\label{sec:framework-overview}

\begin{figure}
    \centering
    \includegraphics[width=\linewidth]{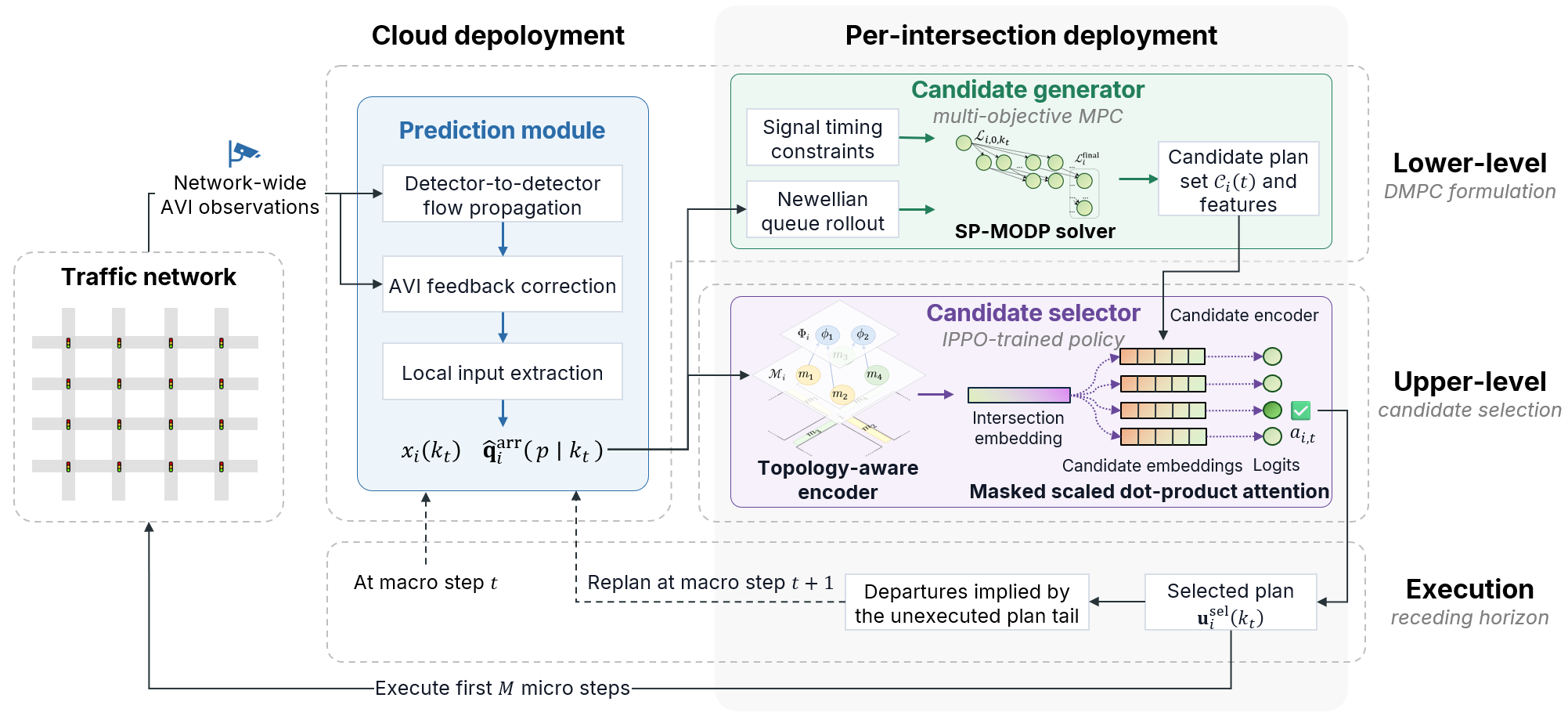}
    \caption{Deployment-aware overview of SelectLight. The cloud-deployed prediction module and per-intersection candidate generators jointly implement the lower-level DMPC formulation, while each candidate selector applies the shared IPPO-trained upper-level policy. At every macro step, the selected plan is executed for $M$ micro steps, departures implied by its unexecuted tail feed the next arrival prediction, and the framework replans.}
    \label{fig:framework-overview}
\end{figure}

Figure~\ref{fig:framework-overview} maps the hierarchical formulation in Section~\ref{sec:problem-formulation} to the deployment architecture of SelectLight. A cloud-deployed prediction module handles network-wide information, while each intersection contains a candidate generator and a candidate selector. The prediction module and candidate generators jointly implement the lower-level DMPC formulation, whereas the candidate selectors apply a shared upper-level policy. This division separates traffic information processing, constraint-aware signal planning, and adaptive plan selection.

At the beginning of each macro step, the prediction module combines network-wide automatic vehicle identification (AVI) records with departures implied by previously selected upstream plans. A calibrated detector-to-detector model propagates these departures through the network to predict future arrivals, while newly observed AVI records correct accumulated prediction errors. For each intersection $i$, the module provides two inputs: the current local traffic state $x_i(k_t)$ and the lane-level predictive arrival profile $\hat{\mathbf q}_i^{\mathrm{arr}}(p\mid k_t)$. These inputs allow each agent to account for upstream traffic without solving a centralized network-wide signal control problem.

Given these inputs, the local agent first invokes its candidate generator. The Newellian point--spatial queue model predicts the traffic effects of phase-based signal plans, and the signal timing constraints exclude operationally infeasible plans. SP-MODP solves the resulting local multi-objective MPC problem with respect to total queueing delay, peak queue accumulation, and total number of stops. Its output is a bounded set of mutually nondominated feasible candidate plans $\mathcal C_i(t)$, together with the objective values and stage descriptors used to distinguish their trade-offs.

The candidate selector then chooses one plan from $\mathcal C_i(t)$. It represents the current traffic context and the attributes of each candidate plan, and uses masked scaled dot-product attention to score only the available plans. The selector's policy is trained with IPPO using realized local queueing delay as its reward. Consequently, the prediction module and candidate generators handle network coupling, candidate plan evaluation, signal constraints, and objective trade-offs, whereas learning is confined to choosing among plans already found feasible.

Only the first $M$ micro steps of the selected plan are executed. During this interval, AVI devices record realized vehicle passages; meanwhile, departures implied by the unexecuted plan tail contribute to the next arrival prediction. At the following macro step, the prediction module updates the local inputs and each agent generates and selects a new plan. Sections~\ref{sec:traffic-prediction}, \ref{sec:lower-level-dmpc-solver}, and~\ref{sec:upper-level-rl-policy} detail arrival prediction and traffic state estimation, multi-objective MPC candidate generation, and attention-based candidate selection, respectively.

\section{Arrival prediction and traffic state estimation}
\label{sec:traffic-prediction}

Online candidate generation and selection require a current local traffic state and a lane-level predictive arrival profile for each intersection. The predictive profile carries upstream traffic information and couples the local MPC problems. The prediction module constructs both inputs from network-wide AVI observations and departures implied by previously selected plans. It propagates these departures through a calibrated detector-to-detector model, corrects accumulated errors with subsequent AVI observations, and extracts the local inputs supplied to the candidate generators and candidate selectors.

\subsection{AVI-based arrival prediction}
\label{sec:avi-deployment-arrival-prediction}

\begin{figure}
  \centering
  \includegraphics[width=0.9\linewidth]{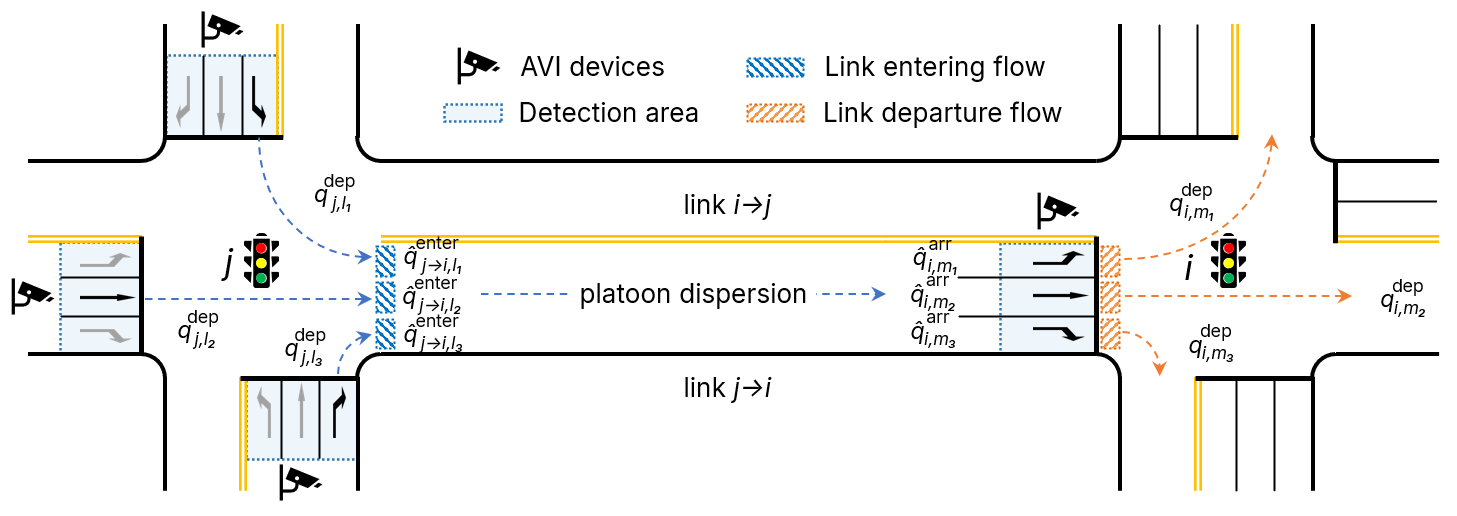}
  \caption{AVI sensing configuration and traffic propagation between adjacent signalized intersections. Devices near the stop lines record vehicle identities and passage times, while their detection areas provide local evidence of whether vehicles are present near the stop line.}
  \label{fig:avi-deployment}
\end{figure}

Figure~\ref{fig:avi-deployment} illustrates the AVI sensing configuration. Devices are installed immediately upstream of the stop lines and associate each detected vehicle with its lane and passage time. Following \citet{luo2025probabilistic}, the effective detection area is assumed to extend approximately 0--20~m upstream of the stop line, covering the first few vehicles near the stop line. The resulting records support both vehicle re-identification between adjacent intersections and detection of a persistently empty stop-line region. The lane-level quantities in the figure distinguish observed departure flows $q_{j,l}^{\mathrm{dep}}$ and $q_{i,m}^{\mathrm{dep}}$, predicted link entering flows $\hat q_{j\to i,l}^{\mathrm{enter}}$, and predicted downstream arrivals $\hat q_{i,m}^{\mathrm{arr}}$, where $l$ and $m$ index lanes at intersections $j$ and $i$, respectively. Time arguments are omitted in the figure for readability; the following formulation stacks the corresponding quantities over all detectors.

Departures observed at an upstream intersection are mapped to downstream arrivals through a calibrated platoon dispersion model. The model combines three physical effects: the travel time offset from the upstream stop line into the connecting link, the distribution of link travel times between adjacent intersections, and the lane-to-lane turning fractions. Its detailed calibration follows \citet{luo2025distributed}.

For network-wide computation, these effects are jointly encoded by a stationary flow-propagation tensor $\mathcal W\in\mathbb R^{n_{\mathrm d}\times T^{\mathrm{disp}}\times n_{\mathrm d}}$, where $n_{\mathrm d}$ is the number of AVI detectors and $T^{\mathrm{disp}}$ is the maximum propagation horizon. The element $\mathcal W_{d,\tau,d'}$ is the expected fraction of flow observed at detector $d$ that reaches detector $d'$ after $\tau$ micro steps. Let $\mathbf W^{(\tau)}$ denote the temporal slice of $\mathcal W$ at lag $\tau$. At micro step $k_t$, the onset of macro step $t$, the network arrival profile is computed as
\begin{equation}
  \hat{\mathbf q}^{\mathrm{arr}}(k_t+p\mid k_t)
  =
  \sum_{\tau=0}^{T^{\mathrm{disp}}-1}
  \left(\mathbf W^{(\tau)}\right)^\top
  \tilde{\mathbf q}^{\mathrm{dep}}(k_t+p-\tau\mid k_t),
  \qquad
  p=0,\dots,H-1,
  \label{eq:network-arrival-prediction}
\end{equation}
where $\tilde{\mathbf q}^{\mathrm{dep}}(k_t+p-\tau\mid k_t)$ equals the measured departure vector $\mathbf q^{\mathrm{dep}}(k_t+p-\tau)$ when $p-\tau<0$. When $p-\tau\geq0$, it combines boundary departures forecast from recent AVI measurements with the departures implied by the signal plans committed at the preceding macro step. Equation~\eqref{eq:network-arrival-prediction} can therefore be evaluated by batched matrix operations while retaining lane- and time-specific arrival information.

\subsection{Cumulative flow estimation with AVI feedback}

Propagation alone can accumulate state errors when realized travel times or vehicle routes differ from their calibrated distributions. The AVI records provide two complementary feedback signals that correct these errors without introducing a separate stochastic state estimator.

The first signal is downstream vehicle re-identification. After a vehicle is observed at detector $d$, its possible downstream arrival times and lanes are represented by the corresponding entries of $\mathcal W$. When the same vehicle is subsequently identified at detector $d'$, its realized destination and passage time replace the unresolved contribution of the earlier prediction, and a new prediction is initiated from detector $d'$. This vehicle-level update prevents an already observed vehicle from continuing to contribute to an obsolete downstream arrival or queue prediction.

The second signal is persistent absence of vehicles from an internal AVI detection area. If the detection area of a lane remains empty after the arrival times assigned by the predictions have elapsed, the observation provides negative evidence against the predicted accumulation. The inconsistent expected vehicle count is removed from that lane and redistributed over the other feasible downstream lanes of the same upstream stream by renormalizing their calibrated turning fractions. If no destination remains consistent with the observations, that expected count is excluded from the current queue estimate.

To express the resulting cumulative flow state, let $\mathbf B(k)$ denote the unresolved expected vehicle count that has propagated to the network detectors by micro step $k$ but has not yet been resolved by downstream AVI observations. The measured cumulative departures, estimated cumulative arrivals, and network point queue satisfy
\begin{equation}
\begin{aligned}
  \mathbf N^{\mathrm{dep}}(k)
  &=
  \mathbf N^{\mathrm{dep}}(k-1)+\mathbf q^{\mathrm{dep}}(k),
  \\
  \mathbf N^{\mathrm{arr}}(k)
  &=
  \mathbf N^{\mathrm{dep}}(k)+\mathbf B(k),
  \\
  \mathbf Q^{\mathrm p}(k)
  &=
  \mathbf N^{\mathrm{arr}}(k)-\mathbf N^{\mathrm{dep}}(k)
  =\mathbf B(k)\geq\mathbf 0.
\end{aligned}
\label{eq:feedback-corrected-cumulative-flow}
\end{equation}
Vehicle re-identification and empty detection area feedback update $\mathbf B(k)$ as new AVI records arrive. Thus, the point queue remains consistent with the cumulative counts and nonnegative by construction.

\subsection{Local traffic state and predictive arrival profile}

The cumulative counts determine the point queue but do not record the progress of an active queue dissipation front. The prediction module therefore propagates the auxiliary front state $\boldsymbol\zeta$ under the executed segment of the selected plan using the Newellian dynamics in Section~\ref{sec:lower-queue-models}. At each macro step, the feedback-corrected cumulative counts and the propagated front state form the initial condition for the next queue rollout.

Let $\mathbf S_i$ be the binary matrix that selects the inbound lanes of intersection $i$ from the network-level detector vectors. The local state supplied at micro step $k_t$, the onset of macro step $t$, is
\begin{equation}
  x_i(k_t)
  =
  \left(
  \mathbf S_i\mathbf N^{\mathrm{arr}}(k_t),
  \;
  \mathbf S_i\mathbf N^{\mathrm{dep}}(k_t),
  \;
  \boldsymbol\zeta_i(k_t)
  \right),
  \label{eq:local-initial-state}
\end{equation}
where $\boldsymbol\zeta_i(k_t)=\{\zeta_{i,m}(k_t)\}_{m\in\mathcal M_i}$ contains the lane-level queue dissipation front states. The point and spatial queues are derived from $x_i(k_t)$ through Equations~\eqref{eq:point-queue-dynamics} and~\eqref{eq:spatial-queue-dynamics}, respectively. The corresponding predictive arrival profile is
\begin{equation}
  \hat{\mathbf q}_i^{\mathrm{arr}}(p\mid k_t)
  =
  \mathbf S_i
  \hat{\mathbf q}^{\mathrm{arr}}(k_t+p\mid k_t),
  \qquad
  p=0,\dots,H-1.
  \label{eq:local-arrival-extraction}
\end{equation}

The profile in Equation~\eqref{eq:local-arrival-extraction} aggregates flows propagated from departures observed before $k_t$, forecast boundary departures, and departures implied by previously selected upstream plans. Together, Equations~\eqref{eq:network-arrival-prediction} and~\eqref{eq:local-arrival-extraction} instantiate the coupling operator $\mathcal{H}_i^{\mathrm{arr}}$ introduced in Equation~\eqref{eq:dmpc-coupling}. The resulting lane-level profile is used by the SP-MODP solver within each candidate generator to evaluate candidate signal plans and by each candidate selector to represent predicted traffic conditions.

\section{Multi-objective MPC candidate generation}
\label{sec:lower-level-dmpc-solver}

Post-optimization selection requires the lower level to construct several feasible plans with distinct predicted trade-offs at every control update. At macro step $t$, the candidate generator receives the local state $x_i(k_t)$ and predictive arrival profile $\hat{\mathbf q}_i^{\mathrm{arr}}(p\mid k_t)$ from the prediction module (Section~\ref{sec:traffic-prediction}). Directly retaining all nondominated partial plans would make the online search difficult to bound. SP-MODP therefore applies objective space pruning and prescribed caps to return a bounded set of mutually nondominated feasible candidate signal plans $\mathcal C_i(t)$. The candidate selector chooses one of these plans for execution.

\subsection{Multi-objective problem formulation}
\label{sec:lower-objective}

The candidate generator evaluates each candidate signal plan from three complementary operational perspectives:
\begin{enumerate}[label=(\arabic*), leftmargin=*]
    \item \textbf{Total queueing delay}. The cumulative time that vehicles are predicted to spend in standing queues over the prediction horizon.

    \item \textbf{Peak queue accumulation}. The worst predicted aggregate queue condition over the prediction horizon. This objective gives greater weight to localized queue accumulation associated with spillback risk.

    \item \textbf{Total number of stops}. The aggregate number of initial and repeated stop events estimated over the prediction horizon. This objective characterizes interruptions to vehicle progression.
\end{enumerate}

Delay, queue length, and number of stops are widely used measures of effectiveness in traffic signal control, but they characterize different aspects of traffic operations \citep{qadri2020stateofart,jiao2016pareto}. Total delay reflects aggregate traffic efficiency, peak queue accumulation penalizes localized congestion, and the number of stops characterizes interrupted vehicle progression. Scalarizing these criteria requires their relative priorities to be specified before optimization \citep{jiao2016pareto,alislam2017distributed}. The candidate generator instead retains a bounded, mutually nondominated candidate set \citep{li2019multiobjective,luo2025distributed}, from which the candidate selector chooses one plan according to the current traffic context.

Formally, the candidate set at macro step $t$ is drawn from the feasible plan space
\begin{equation}
\mathcal C_i(t)
\subset
\left\{
\mathbf u_i
=
\{u_{i,m}(k)\}_{k=k_t}^{k_t+H-1}
:
\mathbf u_i
\text{ satisfies the phase structure constraints}
\right\},
\end{equation}
The ideal search target consists of plans that are Pareto-optimal with respect to
\begin{equation}
\mathbf F_i(k_t)
=
\begin{bmatrix}
F_i^{\mathrm{delay}}(k_t)\\[2pt]
F_i^{\mathrm{queue}}(k_t)\\[2pt]
F_i^{\mathrm{stop}}(k_t)
\end{bmatrix}.
\label{eq:objective-vector}
\end{equation}
The bounded pruning in SP-MODP approximates this target while preserving feasibility, as detailed in Section~\ref{sec:lower-spmodp}. The queue model and the precise definitions of these objectives are given next.

\subsection{Newellian point--spatial queue model}
\label{sec:lower-queue-models}

\begin{figure}
    \centering
    \includegraphics[width=\linewidth]{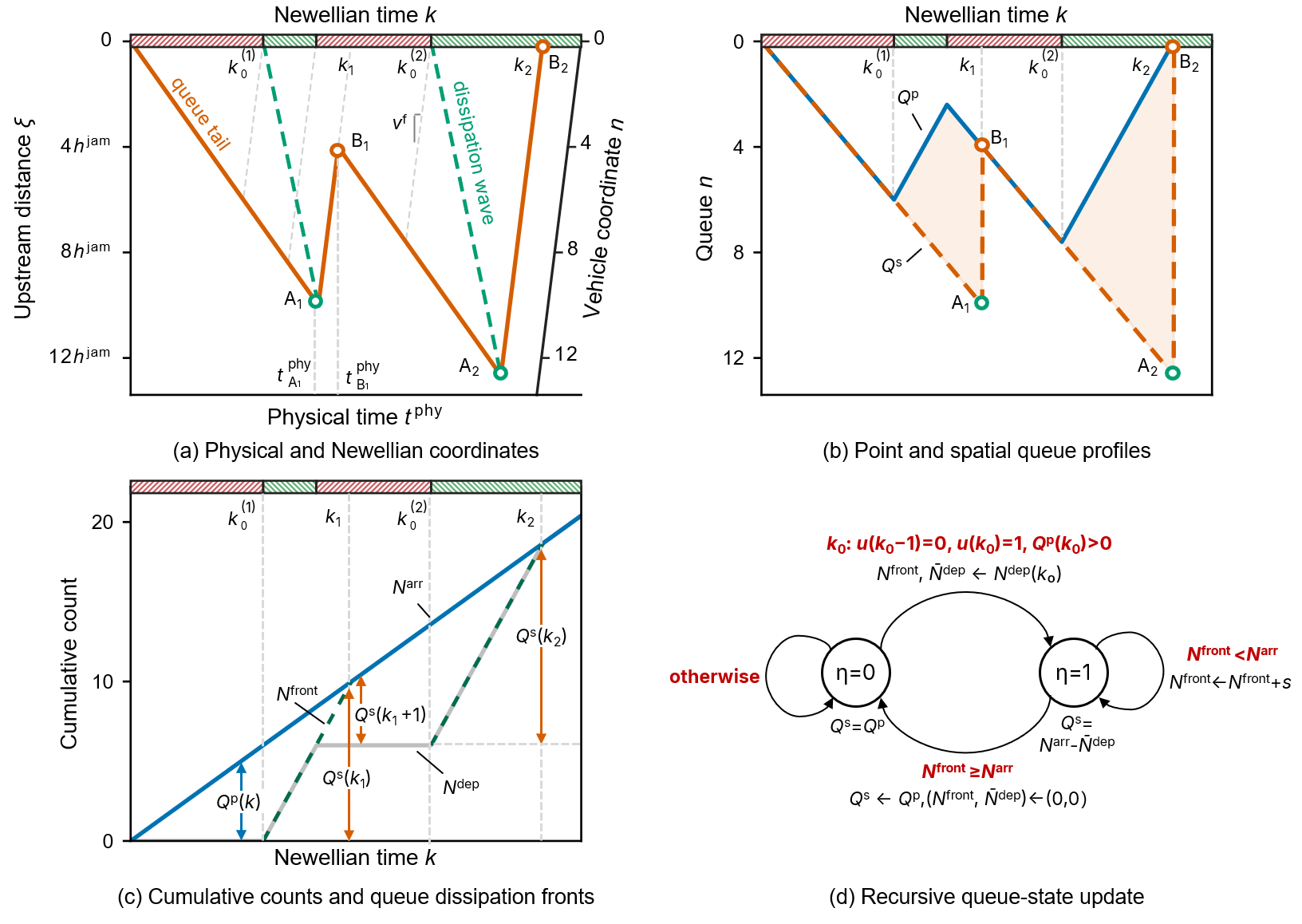}
    \caption{Illustration of the deterministic Newellian point--spatial queue model. (a) Physical and Newellian coordinates. (b) Point and spatial queue profiles. (c) Cumulative counts and queue dissipation fronts. (d) Recursive queue state update.}
    \label{fig:newellian-queue-model}
\end{figure}

We evaluate each candidate signal plan using a deterministic point--spatial queue model under Newellian coordinates \citep{newell2002simplified,wang2024traffic}. The platoon dispersion model in Section~\ref{sec:traffic-prediction} accounts for heterogeneous link travel times and produces the lane-level predictive arrival profile used by the candidate generator. Given this profile, the local queue rollout assumes first-in, first-out service within each lane and uses lane-specific effective values of free-flow speed and jam spacing. Thus, the homogeneity assumption applies to the within-lane queue representation rather than to upstream traffic propagation. Figure~\ref{fig:newellian-queue-model} summarizes the coordinate transformation, the two queue representations, their cumulative count interpretation, and the recursive queue state update.

Figure~\ref{fig:newellian-queue-model}(a) illustrates the mapping from physical coordinates to Newellian coordinates. Consider an event on lane $m$ of agent $i$ at physical time $t^{\mathrm{phy}}$ and upstream distance $\xi\geq 0$. Its Newellian coordinates $(k,n)$ are
\begin{equation}
k
=
\frac{t^{\mathrm{phy}}}{\Delta_k}
+\frac{\xi}{v_{i,m}^{\mathrm f}\Delta_k},
\qquad
n
=
\frac{\xi}{h_{i,m}^{\mathrm{jam}}},
\label{eq:newellian-coordinate}
\end{equation}
where $\Delta_k$ is the micro-step duration, $v_{i,m}^{\mathrm f}$ is the free-flow speed, and $h_{i,m}^{\mathrm{jam}}$ is the jam spacing. Thus, $k$ is the stop line arrival index that the event would attain under uninterrupted free-flow travel, whereas $n$ is its vehicle coordinate measured in jam spacings. At the stop line, $k=t^{\mathrm{phy}}/\Delta_k$, so the queue model and signal plan share the same discrete time grid.

As shown in Figure~\ref{fig:newellian-queue-model}(b), the model retains two complementary queue states. The \emph{point queue} $Q_{i,m}^{\mathrm p}(k)$ is the number of stopped vehicle equivalents, whereas the \emph{spatial queue} $Q_{i,m}^{\mathrm s}(k)$ is the vehicle-equivalent coordinate of the last stopped vehicle. The corresponding physical queue length is
\begin{equation}
L_{i,m}^{\mathrm q}(k)
=
h_{i,m}^{\mathrm{jam}}Q_{i,m}^{\mathrm s}(k).
\label{eq:spatial-queue-distance}
\end{equation}
Under the inverse coordinate mapping, the queue-tail state $(k,Q_{i,m}^{\mathrm s}(k))$ occurs at physical time $t^{\mathrm{phy}}=k\Delta_k-L_{i,m}^{\mathrm q}(k)/v_{i,m}^{\mathrm f}$. Both queue states are measured in vehicle equivalents and may be fractional because the predicted arrivals are expected counts. They satisfy
\begin{equation}
0
\leq
Q_{i,m}^{\mathrm p}(k)
\leq
Q_{i,m}^{\mathrm s}(k),
\qquad
Q_{i,m}^{\mathrm p}(k)=0
\Longleftrightarrow
Q_{i,m}^{\mathrm s}(k)=0.
\label{eq:point-spatial-feasibility}
\end{equation}
The distinction matters during queue discharge: departures immediately reduce $Q_{i,m}^{\mathrm p}$, whereas $Q_{i,m}^{\mathrm s}$ remains unchanged until the physical queue dissipation wave reaches the queue tail. The shaded regions in Figure~\ref{fig:newellian-queue-model}(b) show intervals in which $Q_{i,m}^{\mathrm s}>Q_{i,m}^{\mathrm p}$.

Figure~\ref{fig:newellian-queue-model}(c) gives the cumulative count interpretation of the queue states. Given a candidate signal plan, let
\begin{equation}
a_{i,m}(k)
\triangleq
\hat q_{i,m}^{\mathrm{arr}}(k-k_t\mid k_t)
\label{eq:local-arrival-shorthand}
\end{equation}
denote the expected arrivals during Newellian time step $k$. This quantity need not be an integer because the traffic prediction model distributes upstream departures probabilistically over downstream arrival times and lanes. Let $u_{i,m}(k)\in\{0,1\}$ indicate whether the lane has right-of-way, and let $s_{i,m}$ be its saturation discharge capacity per micro step. Applying arrivals before departures gives
\begin{equation}
N_{i,m}^{\mathrm{arr}}(k+1)
=
N_{i,m}^{\mathrm{arr}}(k)+a_{i,m}(k),
\label{eq:cumulative-arrival-dynamics}
\end{equation}
\begin{equation}
N_{i,m}^{\mathrm{dep}}(k+1)
=
\min\!\left\{
N_{i,m}^{\mathrm{arr}}(k+1),
\;
N_{i,m}^{\mathrm{dep}}(k)+u_{i,m}(k)s_{i,m}
\right\}.
\label{eq:cumulative-departure-dynamics}
\end{equation}
The point queue is the vertical gap between $N^{\mathrm{arr}}$ and $N^{\mathrm{dep}}$ in Figure~\ref{fig:newellian-queue-model}(c) and follows directly from conservation:
\begin{equation}
Q_{i,m}^{\mathrm p}(k)
=
N_{i,m}^{\mathrm{arr}}(k)-N_{i,m}^{\mathrm{dep}}(k).
\label{eq:point-queue-dynamics}
\end{equation}

The spatial queue additionally depends on the physical queue dissipation wave shown in Figure~\ref{fig:newellian-queue-model}(a). We represent the same moving boundary on the cumulative count axis in Figure~\ref{fig:newellian-queue-model}(c) as a queue dissipation front. Let $\eta_{i,m}(k)\in\{0,1\}$ be the active front indicator, with $\eta_{i,m}(k)=1$ while a front is being propagated. Let $N_{i,m}^{\mathrm{front}}(k)$ denote the current front position, and let $\bar N_{i,m}^{\mathrm{dep}}(k)$ denote the cumulative departure count stored at the onset of the current dissipation episode. A service onset occurs at $k_0$ when $u_{i,m}(k_0-1)=0$ and $u_{i,m}(k_0)=1$. If no front is active and $Q_{i,m}^{\mathrm p}(k_0)>0$, the service onset activates a front and initializes $\eta_{i,m}(k_0)=1$ and $N_{i,m}^{\mathrm{front}}(k_0)=\bar N_{i,m}^{\mathrm{dep}}(k_0)=N_{i,m}^{\mathrm{dep}}(k_0)$. The stored departure count remains unchanged during this dissipation episode, and the front position evolves as
\begin{equation}
N_{i,m}^{\mathrm{front}}(k)
=
\bar N_{i,m}^{\mathrm{dep}}(k_0)
+s_{i,m}(k-k_0),
\qquad k\geq k_0.
\label{eq:dissipation-front-position}
\end{equation}
The green dashed lines in Figure~\ref{fig:newellian-queue-model}(c) show how the front position evolves at the saturation rate. Both $N_{i,m}^{\mathrm{front}}(k)$ and $N_{i,m}^{\mathrm{arr}}(k)$ are expressed on the cumulative vehicle-count axis: the former identifies the vehicle coordinate reached by the front, whereas the latter identifies the queue-tail coordinate. Therefore, the front has not yet reached the queue tail and continues to propagate while
\begin{equation}
N_{i,m}^{\mathrm{front}}(k)<N_{i,m}^{\mathrm{arr}}(k).
\label{eq:dissipation-front-active}
\end{equation}
Once $N_{i,m}^{\mathrm{front}}(k)\geq N_{i,m}^{\mathrm{arr}}(k)$, the front has reached the queue tail. The post-event value is recorded in the next discrete state by deactivating the front and resetting the spatial queue: $\eta_{i,m}(k+1)=0$ and $Q_{i,m}^{\mathrm s}(k+1)=Q_{i,m}^{\mathrm p}(k+1)$. At most one front is active on a lane at a time. The spatial queue is therefore
\begin{equation}
Q_{i,m}^{\mathrm s}(k)
=
\begin{cases}
N_{i,m}^{\mathrm{arr}}(k)-\bar N_{i,m}^{\mathrm{dep}}(k),
& \eta_{i,m}(k)=1,
\\[4pt]
Q_{i,m}^{\mathrm p}(k),
& \eta_{i,m}(k)=0.
\end{cases}
\label{eq:spatial-queue-dynamics}
\end{equation}
Figure~\ref{fig:newellian-queue-model}(a)--(c) illustrates two dissipation episodes. At $A_j$, the front reaches the queue tail, i.e., $N^{\mathrm{front}}=N^{\mathrm{arr}}$, and the spatial queue is reset instantaneously to the point queue at $B_j$. In Figure~\ref{fig:newellian-queue-model}(c), $k_1$ and $k_1+1$ illustrate how the states immediately before and after this reset are recorded on the discrete grid; they do not imply a finite reset duration. The first reset leaves a positive residual queue, whereas the second clears the queue because $Q^{\mathrm p}=0$.

Figure~\ref{fig:newellian-queue-model}(d) summarizes this recursive update. The two queue values alone are not a sufficient recursive state: they do not determine whether a front is active, how far it has propagated, or which cumulative departure count is stored for the current dissipation episode. Consequently, two states with identical $(Q^{\mathrm p},Q^{\mathrm s})$ can evolve differently under the same future arrivals and signal plan. We therefore retain the following lane-level auxiliary state alongside the cumulative arrival and departure counts:
\begin{equation}
\zeta_{i,m}(k)
=
\left(
\eta_{i,m}(k),
N_{i,m}^{\mathrm{front}}(k),
\bar N_{i,m}^{\mathrm{dep}}(k)
\right),
\label{eq:dissipation-front-state}
\end{equation}
where $N_{i,m}^{\mathrm{front}}(k)$ and $\bar N_{i,m}^{\mathrm{dep}}(k)$ are set to zero when $\eta_{i,m}(k)=0$. The onset time need not be stored because the current front position and stored cumulative departure count are sufficient for subsequent propagation. This richer recursive state must be retained for each partial signal plan and therefore directly shapes the SP-MODP construction in Section~\ref{sec:lower-spmodp}.

\subsection{Objective functions}
\label{sec:lower-objective-computation}

\paragraph{Total queueing delay}
Consistent with the input--output accounting identity for traffic delay~\citep{daganzo1983derivation}, the predicted queueing delay is computed as the discrete-time area under the lane-level point-queue trajectories:
\begin{equation}
F_i^{\mathrm{delay}}(k_t)
=
\sum_{k=k_t+1}^{k_t+H}
\sum_{m\in\mathcal M_i}
Q_{i,m}^{\mathrm p}(k).
\label{eq:objective-delay}
\end{equation}
The result is measured in vehicle-micro-steps. Since $\Delta_k=1$~s in the experiments, it is numerically equivalent to vehicle-seconds.

\paragraph{Peak queue accumulation}
The worst aggregate queue condition over the prediction horizon is computed using the maximum quadratic aggregation of lane-level spatial queues:
\begin{equation}
F_i^{\mathrm{queue}}(k_t)
=
\max_{k\in\{k_t+1,\dots,k_t+H\}}
\sum_{m\in\mathcal M_i}
\left(
Q_{i,m}^{\mathrm s}(k)
\right)^2.
\label{eq:objective-queue}
\end{equation}
Here, accumulation denotes aggregation across lanes at a given micro step, not summation over time. The quadratic aggregation gives greater weight to long queues on individual lanes, while the maximum operator captures the worst aggregated queue condition over the prediction horizon.

\paragraph{Total number of stops}
For each lane $m$ and micro step $k\in\{k_t,\dots,k_t+H-1\}$, the aggregate incremental stop count estimated from the queue rollout is
\begin{equation}
\psi_{i,m}(k)
=
\mathbf{1}\!\left\{
Q_{i,m}^{\mathrm s}(k+1)>0
\right\}a_{i,m}(k)
+
\mathbf{1}\!\left\{
\eta_{i,m}(k)=1
\land
\eta_{i,m}(k+1)=0
\right\}Q_{i,m}^{\mathrm s}(k+1),
\label{eq:objective-stop-increment}
\end{equation}
where $\mathbf{1}\{\cdot\}$ is the indicator function. The total number of stops is
\begin{equation}
F_i^{\mathrm{stop}}(k_t)
=
\sum_{k=k_t}^{k_t+H-1}
\sum_{m\in\mathcal M_i}
\psi_{i,m}(k).
\label{eq:objective-stop}
\end{equation}
The first term counts the initial stop of arrivals that encounter a standing queue. The second term approximates repeated stops by counting the residual queue whenever the dissipation front reaches the queue tail. If this occurs over multiple signal cycles, the same vehicles may be counted more than once. Because the rollout operates on expected arrivals and aggregate queue states, this objective is a queue-based estimate rather than a reconstruction of individual vehicle trajectories; exact realized stop counts are evaluated separately from the microscopic simulation.

\subsection{Phase-based decision-space reformulation}
\label{sec:lower-phase}

The phase-based reformulation serves a computational role by converting the micro-step control problem into the finite stage--time lattice required by SP-MODP. The unconstrained representation $\{u_{i,m}(k)\}$ contains $H|\mathcal M_i|$ binary variables and admits up to $2^{H|\mathcal M_i|}$ right-of-way sequences. For the cyclic phase structure developed below, compatible movement phases follow a prescribed order and satisfy green-time and intergreen requirements. We therefore replace the binary sequence with a bounded sequence of integer-valued phase end times. Under a fixed phase order, these times uniquely determine the phase durations and micro-step right-of-way sequence.

\begin{figure}
    \centering
    \includegraphics[width=\linewidth]{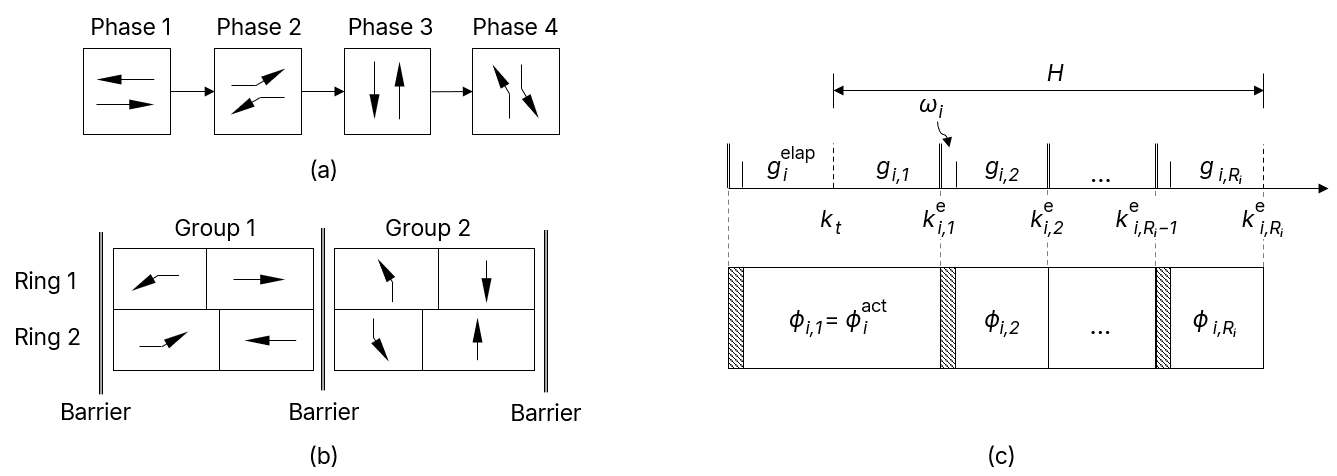}
    \caption{Phase structures and phase-based horizon representation. (a) Standard four-phase structure used as the running example. (b) Ring-and-barrier structure shown as a possible extension. (c) Horizon representation at a macro step with no residual intergreen.}
    \label{fig:phase-reformulation}
\end{figure}

Figure~\ref{fig:phase-reformulation}(a) shows the four-phase structure used as the running example. Figure~\ref{fig:phase-reformulation}(b) illustrates a possible extension to ring-and-barrier control. Each barrier group can be treated as one stage block, while its feasible internal phase combinations can be enumerated following \citet{feng2015realtime}. Figure~\ref{fig:phase-reformulation}(c) shows the cyclic horizon representation developed below.

Let $\Phi_i$ be the feasible phase set, with $\operatorname{succ}_i(\cdot)$ and $\operatorname{pred}_i(\cdot)$ defining its cyclic order. At macro step $t$, $\phi_i^{\mathrm{act}}$ denotes the active green phase or the phase scheduled after an ongoing intergreen. The remaining intergreen is $y_i^{\mathrm{rem}}\in\mathbb Z_{\geq 0}$, and subsequent switches use the full intergreen $\omega_i$. The elapsed green $g_i^{\mathrm{elap}}$ applies only when $y_i^{\mathrm{rem}}=0$ and is otherwise zero. All timing quantities below are measured in micro steps.

A feasible plan contains a plan-dependent number of stages $R_i\in\{1,\dots,\bar R\}$. Stage $r$ has phase $\phi_{i,r}\in\Phi_i$, within-horizon green duration $g_{i,r}\in\mathbb Z_{\geq0}$, and green end time $k_{i,r}^{\mathrm e}$. The end-time sequence $\mathbf{k}_i^{\mathrm e}=(k_{i,1}^{\mathrm e},\dots,k_{i,R_i}^{\mathrm e})$ satisfies
\begin{alignat}{2}
\phi_{i,1}&=\phi_i^{\mathrm{act}},
\qquad \phi_{i,r+1}=\operatorname{succ}_i(\phi_{i,r}),
&&\qquad r=1,\dots,R_i-1,
\label{eq:phase-order}\\
k_{i,1}^{\mathrm e}&=k_t+y_i^{\mathrm{rem}}+g_{i,1},
\label{eq:first-phase-end}\\
k_{i,r+1}^{\mathrm e}&=k_{i,r}^{\mathrm e}+\omega_i+g_{i,r+1},
&&\qquad r=1,\dots,R_i-1,
\label{eq:phase-end-recursion}\\
k_{i,R_i}^{\mathrm e}&=k_t+H.
\label{eq:phase-horizon-closure}
\end{alignat}
The final equality truncates the last stage at the prediction-horizon boundary. Figure~\ref{fig:phase-reformulation}(c) illustrates the case $y_i^{\mathrm{rem}}=0$.

Let $\mathcal M_i(\phi)\subseteq\mathcal M_i$ contain the lanes served by phase $\phi$. Lanes served by both adjacent phases retain right-of-way during their transition, so $\mathcal M_i^{\mathrm{tr}}(\phi,\phi')=\mathcal M_i(\phi)\cap\mathcal M_i(\phi')$. With $\phi_{i,0}=\operatorname{pred}_i(\phi_{i,1})$, the green and intergreen intervals are
\begin{alignat*}{3}
\mathcal K_{i,1}^{\mathrm g}
&=\{k_t+y_i^{\mathrm{rem}},\dots,k_{i,1}^{\mathrm e}-1\}, &\qquad
\mathcal K_{i,r}^{\mathrm g}
&=\{k_{i,r-1}^{\mathrm e}+\omega_i,\dots,k_{i,r}^{\mathrm e}-1\}, &&\qquad r\geq2,\\
\mathcal K_{i,0}^{\mathrm{ig}}
&=\{k_t,\dots,k_t+y_i^{\mathrm{rem}}-1\}, &\qquad
\mathcal K_{i,r}^{\mathrm{ig}}
&=\{k_{i,r}^{\mathrm e},\dots,k_{i,r}^{\mathrm e}+\omega_i-1\}, &&\qquad r=1,\dots,R_i-1.
\end{alignat*}
The initial intergreen set is empty when $y_i^{\mathrm{rem}}=0$. For lane $m\in\mathcal M_i$ and micro step $k\in\{k_t,\dots,k_t+H-1\}$,
\begin{equation}
u_{i,m}(k)=1
\quad\Longleftrightarrow\quad
\begin{aligned}
&\exists r\in\{1,\dots,R_i\}:\;
m\in\mathcal M_i(\phi_{i,r}),\;
k\in\mathcal K_{i,r}^{\mathrm g},
\\[-2pt]
&\text{or}\quad
\exists r\in\{0,\dots,R_i-1\}:\;
m\in\mathcal M_i^{\mathrm{tr}}(\phi_{i,r},\phi_{i,r+1}),\;
k\in\mathcal K_{i,r}^{\mathrm{ig}}.
\end{aligned}
\label{eq:phase-induced-control}
\end{equation}
Thus, the phase end sequence uniquely determines the right-of-way sequence, queue trajectories, and objective vector without changing the underlying traffic model.

Let $g_i^{\min}(\phi)$ and $g_i^{\max}(\phi)$ denote the minimum and maximum green times. The feasible durations are
\begin{alignat}{2}
\max\!\left\{0,\;g_i^{\min}(\phi_{i,1})-g_i^{\mathrm{elap}}\right\}
&\le
g_{i,1}
\le
g_i^{\max}(\phi_{i,1})-g_i^{\mathrm{elap}},
\label{eq:phase-green-feasibility-first}
\\
g_i^{\min}(\phi_{i,r})
&\le
g_{i,r}
\le
g_i^{\max}(\phi_{i,r}),
&&\qquad r=2,\dots,R_i-1,
\label{eq:phase-green-feasibility-middle}
\\
1
&\le
g_{i,R_i}
\le
g_i^{\max}(\phi_{i,R_i}),
&&\qquad R_i\geq 2.
\label{eq:phase-green-feasibility-terminal}
\end{alignat}
The terminal stage need not satisfy the usual minimum because it may continue beyond the horizon. When $R_i=1$, Equation~\eqref{eq:phase-green-feasibility-first} applies. Its duration may be zero only if the current phase has already satisfied its minimum green.

We optionally limit abrupt timing changes between receding-horizon updates. Let $\bar{\mathbf{k}}_i^{\mathrm e,\mathrm{ref}}=(\bar{k}_{i,1}^{\mathrm e},\dots,\bar{k}_{i,R_i^{\mathrm{ref}}}^{\mathrm e})$ contain the unexecuted, nonterminal phase boundaries inherited from the previously selected plan. The common ordered prefix satisfies
\begin{equation}
\left|k_{i,r}^{\mathrm e}-\bar{k}_{i,r}^{\mathrm e}\right|
\le
\Delta_i^{\mathrm e},
\qquad
r=1,\dots,\min\{R_i-1,\;R_i^{\mathrm{ref}}\},
\label{eq:phase-interopt-consistency}
\end{equation}
where $\Delta_i^{\mathrm e}$ is a prescribed tolerance. If phase-end times are stored as absolute micro-step indices, the retained reference boundaries require no shift. If they are stored relative to the previous optimization start, $M$ is subtracted from each retained boundary before comparison. These constraints define the finite stage--time lattice searched by SP-MODP.

\subsection{State-pruned multi-objective dynamic programming}
\label{sec:lower-spmodp}
The phase-based reformulation admits a dynamic program similar to the classical controlled optimization of phases (COP) algorithm \citep{sen1997controlled}. Population-based multi-objective evolutionary algorithms are also applicable to this discrete problem \citep{deb2002fast}. However, the fixed phase order and recursive queue dynamics make dynamic programming well suited to reusing partial rollouts and enforcing signal timing feasibility by construction. The related MODP formulation in \citet{luo2025distributed} uses the phase index as the DP stage and the phase end time as its state, associating each retained objective vector with predicted movement-level point queues. With the lane-level point--spatial model used here, however, the search requires a richer traffic rollout state: partial plans reaching the same stage and end time can carry different cumulative departures and queue-dissipation-front states, which in turn affect subsequent transitions. A direct formulation that adds these quantities to the structural node index would create a prohibitively large augmented state space.

\begin{figure}
    \centering
    \includegraphics[width=0.9\linewidth]{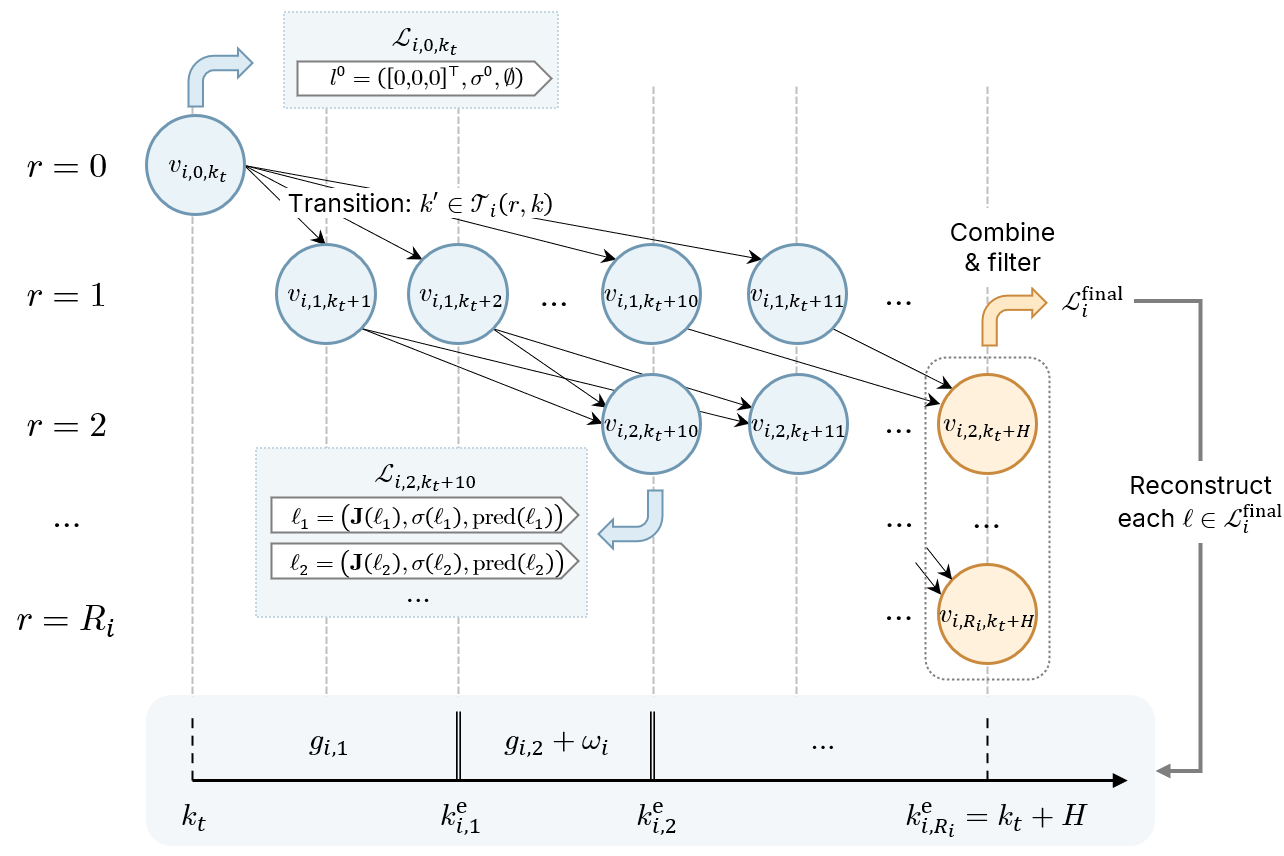}
    \caption{Schematic illustration of SP-MODP on the stage--time lattice. Each node $\nu_{i,r,k}$ represents feasible partial signal plans whose $r$-th stage ends at micro step $k$ and maintains a bounded label set $\mathcal L_{i,r,k}$. Feasible transitions expand labels to successor nodes, after which terminal labels are merged, filtered, and backtracked to reconstruct the retained candidate plans.}
    \label{fig:spmodp-lattice}
\end{figure}

Building on this stage--time organization, SP-MODP avoids augmenting the structural node index with the richer rollout state. Each node remains indexed only by the stage $r$ and phase end time $k$, while the plan-dependent label state $\sigma(\ell)$ is carried within each label. Objective space pruning and the per-node cap $K_i^{\mathrm{pf}}$ bound the state-carrying partial plans propagated from each node. Relative to a fully augmented formulation, this reduces the explicit node state space to the stage--time lattice while retaining the information required for recursive rollout. This combination of reduced node indexing and bounded propagation of state-carrying labels is termed \emph{state-pruned}. Figure~\ref{fig:spmodp-lattice} illustrates this construction, and Algorithm~\ref{alg:spmodp} summarizes the complete label-search procedure.

\paragraph{Stage--time lattice and label structure}

Let $\nu_{i,r,k}$ represent all feasible partial plans whose $r$-th stage ends at micro step $k$. The root is $\nu_{i,0,k_t}$, and any node with $k=k_t+H$ is terminal. Let $\operatorname{Feas}_i(r,k,k')$ indicate that appending phase end time $k'$ satisfies Equations~\eqref{eq:first-phase-end}--\eqref{eq:phase-horizon-closure} and \eqref{eq:phase-green-feasibility-first}--\eqref{eq:phase-interopt-consistency}. The feasible successor set is
\begin{equation}
\mathcal T_i(r,k)
=
\left\{
k'\in\{k,\dots,k_t+H\}:
r<\bar R,
\operatorname{Feas}_i(r,k,k')
\right\},
\label{eq:spmodp-successor-set}
\end{equation}
where $k_{i,0}^{\mathrm e}=k_t$ and $k_{i,r}^{\mathrm e}=k$. Because the phase order is fixed by Equation~\eqref{eq:phase-order}, $k'$ determines the next stage and its micro-step right-of-way sequence.

Each reached node maintains a label set $\mathcal L_{i,r,k}$. Because the cumulative-arrival trajectory is fixed within one optimization, each label stores the plan-dependent cumulative departures together with the auxiliary front state in Equation~\eqref{eq:dissipation-front-state}:
\begin{equation}
\begin{aligned}
\ell
&=
\left(\mathbf J(\ell),\sigma(\ell),\operatorname{pred}(\ell)\right),
\\
\mathbf J(\ell)
&=
\left[J_i^{\mathrm{delay}}(\ell),J_i^{\mathrm{queue}}(\ell),J_i^{\mathrm{stop}}(\ell)\right]^\top,
\\
\sigma(\ell)
&=
\left\{
\left(
N_{i,m}^{\mathrm{dep}}(k_\ell),
\zeta_{i,m}(k_\ell)
\right)
\right\}_{m\in\mathcal M_i},
\end{aligned}
\label{eq:spmodp-label}
\end{equation}
where $k_\ell$ is the phase end time of the attached node, $\mathbf J(\ell)$ is the objective-to-come vector, and $\operatorname{pred}(\ell)$ is a predecessor pointer for path reconstruction. The predictive cumulative arrival trajectory is fixed and indexed by time, so it need not be stored in each label.

\paragraph{Label expansion and objective accumulation}

Let $\mathcal D_i(t)$ collect the predictive arrival profile and queue-model parameters that remain fixed during the current optimization. For a label $\ell\in\mathcal{L}_{i,r,k}$ and a feasible successor $k'\in\mathcal{T}_i(r,k)$, let
\begin{equation}
\bigl(\sigma',\Delta\mathbf{F}_i(\ell;k,k')\bigr)
=
\textsc{Expand}_i\bigl(r,k,k',\sigma(\ell);\mathcal D_i(t)\bigr)
\label{eq:spmodp-expand}
\end{equation}
denote the rollout over $[k,k')$ under the induced stage. Here, $\sigma'=\{(N_{i,m}^{\mathrm{dep}}(k'),\zeta_{i,m}(k'))\}_{m\in\mathcal M_i}$ is the label state reached at $k'$, and $\Delta\mathbf F_i(\ell;k,k')$ contains the corresponding interval contributions to total queueing delay, peak queue accumulation, and total number of stops. Delay and stops accumulate additively, whereas peak queue accumulation propagates through a maximum. We therefore define the stage-wise objective accumulation operator
\begin{equation}
\mathbf{a}\boxplus\mathbf{b}
=
\begin{bmatrix}
a_1+b_1\\
\max\{a_2,b_2\}\\
a_3+b_3
\end{bmatrix},
\qquad
\mathbf{a},\mathbf{b}\in\mathbb{R}^3.
\label{eq:spmodp-boxplus}
\end{equation}
The successor label attached to $\nu_{i,r+1,k'}$ is
\begin{equation}
\ell'
=
\bigl(
\mathbf{J}(\ell)\boxplus\Delta\mathbf{F}_i(\ell;k,k'),\,
\sigma',\,
\ell
\bigr).
\label{eq:spmodp-successor-label}
\end{equation}

\paragraph{Objective-space pruning}

All successor labels reaching the same node are collected in $\widetilde{\mathcal L}_{i,r+1,k'}$. For two labels in this pool, $\ell^a$ dominates $\ell^b$ in objective space if
\begin{equation}
\mathbf{J}(\ell^a)\preceq \mathbf{J}(\ell^b),
\qquad
\mathbf{J}(\ell^a)\neq \mathbf{J}(\ell^b),
\label{eq:spmodp-dominance}
\end{equation}
where $\preceq$ denotes componentwise inequality. Let $\operatorname{Prune}_{K}(\mathcal L)$ discard labels dominated under Equation~\eqref{eq:spmodp-dominance} and, if more than $K$ labels remain, retain the $K$ labels with the largest crowding distances \citep{deb2002fast}. Node updating is therefore $\mathcal L_{i,r+1,k'}=\operatorname{Prune}_{K_i^{\mathrm{pf}}}(\widetilde{\mathcal L}_{i,r+1,k'})$.

SP-MODP uses bounded objective space pruning to balance search coverage and online tractability. Because labels at the same structural node may carry different traffic states, the pruning may not preserve every globally nondominated terminal plan. However, all retained labels remain feasible and are compared jointly in terms of total queueing delay, peak queue accumulation, and total number of stops. Retaining the quadratic queue component during pruning preserves candidates that trade delay and stops against a proxy for future clearance delay and spillback risk.

\paragraph{Terminal processing and candidate reconstruction}

Rather than applying a prescribed minimum-distance rule to return one terminal plan \citep{luo2025distributed}, SP-MODP retains a bounded set for upper-level selection. Terminal labels are merged and reduced to the desired output size:
\begin{equation}
\begin{aligned}
\mathcal L_i^{\mathrm{term}}
&=
\bigcup_{r=1}^{\bar R}\mathcal L_{i,r,k_t+H},
\\
\mathcal L_i^{\mathrm{final}}
&=
\operatorname{Prune}_{K}(\mathcal L_i^{\mathrm{term}}),
\\
\mathcal C_i(t)
&=
\left\{\mathbf u_i(\ell):\ell\in\mathcal L_i^{\mathrm{final}}\right\}.
\end{aligned}
\label{eq:spmodp-output}
\end{equation}
For a terminal label, $\mathbf J(\ell)=\mathbf F_i(k_t;\mathbf u_i(\ell))$. Backtracking its predecessor pointers reconstructs $\mathbf u_i(\ell)$ and the phase end sequence $\mathbf k_i^{\mathrm e}(\ell)$. The phase end sequence and objective vector are retained as descriptors for the candidate selector observation in Section~\ref{sec:upper-level-rl-policy}.

\paragraph{Computational cost}

The structural lattice contains $O(\bar R H)$ nodes, each retaining at most $K_i^{\mathrm{pf}}$ labels. Because each label stores a lane-level traffic state, the label state storage scales as $O(\bar R H K_i^{\mathrm{pf}}|\mathcal M_i|)$. Computation time further depends on the number and duration of feasible transitions and on per-node pruning. These factors are controlled by the signal timing constraints, timing discretization, and label cap. Section~\ref{sec:computational-efficiency} reports the measured solution times under the experimental settings.

\begin{algorithm}[ht!]
\caption{SP-MODP for local candidate generation at macro step $t$}
\label{alg:spmodp}
\DontPrintSemicolon
\SetKwInOut{Input}{Input}
\SetKwInOut{Output}{Output}

\Input{Initial traffic state $\sigma^0$; rollout data $\mathcal D_i(t)$; successor sets $\mathcal T_i(r,k)$; maximum stage count $\bar R$; label cap $K_i^{\mathrm{pf}}$; output size $K$}
\Output{$\mathcal C_i(t)$ and the associated $\mathbf k_i^{\mathrm e}(\ell)$ and $\mathbf J(\ell)$}

Set $\mathcal L_{i,0,k_t}\gets\{([0,0,0]^\top,\sigma^0,\varnothing)\}$\;
\For{$r\gets 0$ \KwTo $\bar R-1$}{
    Clear all incoming pools $\widetilde{\mathcal L}_{i,r+1,k'}$\;
    \ForEach{reached $\nu_{i,r,k}$ with $k<k_t+H$}{
        \ForEach{$\ell\in\mathcal L_{i,r,k}$ and $k'\in\mathcal T_i(r,k)$}{
            $(\sigma',\Delta\mathbf F_i)\gets\textsc{Expand}_i(r,k,k',\sigma(\ell);\mathcal D_i(t))$\;
            Insert $(\mathbf J(\ell)\boxplus\Delta\mathbf F_i,\sigma',\ell)$ into $\widetilde{\mathcal L}_{i,r+1,k'}$\;
        }
    }
    \ForEach{reached $k'$}{
        $\mathcal L_{i,r+1,k'}\gets\operatorname{Prune}_{K_i^{\mathrm{pf}}}(\widetilde{\mathcal L}_{i,r+1,k'})$\;
    }
}
Set $\mathcal L_i^{\mathrm{term}}\gets\bigcup_{r=1}^{\bar R}\mathcal L_{i,r,k_t+H}$\;
Set $\mathcal L_i^{\mathrm{final}}\gets\operatorname{Prune}_{K}(\mathcal L_i^{\mathrm{term}})$\;
Backtrack every $\ell\in\mathcal L_i^{\mathrm{final}}$ and return $\mathcal C_i(t)$ with $\mathbf k_i^{\mathrm e}(\ell)$ and $\mathbf J(\ell)$\;
\end{algorithm}

\section{Attention-based candidate selection}
\label{sec:upper-level-rl-policy}

At each macro step $t$, the candidate generator provides agent $i$ with a bounded set of feasible candidate signal plans $\mathcal C_i(t)$. Because the size and content of this set vary with the traffic state, fixed candidate indices have no persistent meaning across time or intersections. The selector must instead compare the available candidate plan features under current and predicted traffic conditions while accommodating different lane and phase structures. We therefore organize the local observation by physical entity, encode each intersection with topology-aware encoders shared across agents, and score its candidate plans through masked scaled dot-product attention.

\subsection{Observation}
\label{sec:upper-observation-construction}

The observation separates traffic context from candidate plan information. Using the feasible local phase set $\Phi_i$ and the served-lane sets $\mathcal M_i(\phi)$ introduced in Section~\ref{sec:lower-phase}, we represent the physical service relation at intersection $i$ by the bipartite graph
\begin{equation}
\mathcal G_i^{\mathrm{lp}}
=
\left(\mathcal M_i,\Phi_i,\mathcal E_i^{\mathrm{lp}}\right),
\qquad
\mathcal E_i^{\mathrm{lp}}
=
\left\{(m,\phi):m\in\mathcal M_i(\phi)\right\},
\label{eq:upper-lane-phase-graph}
\end{equation}
where an edge indicates that phase $\phi$ serves inbound lane $m$.
The local observation is grouped as
\begin{equation}
o_{i,t}
=
\left(
\mathbf X_{i,t}^{\mathrm{lane}},
\mathbf X_{i,t}^{\mathrm{phase}},
\mathbf x_{i,t}^{\mathrm{int}},
\mathbf X_{i,t}^{\mathrm{cand}},
\mathbf U_{i,t}^{\mathrm{stage}},
\mathbf M_{i,t}^{\mathrm{stage}},
\boldsymbol\mu_{i,t}
\,;\,
\mathcal G_i^{\mathrm{lp}}
\right),
\label{eq:upper-observation}
\end{equation}
Here, $\mathbf X_{i,t}^{\mathrm{lane}}$, $\mathbf X_{i,t}^{\mathrm{phase}}$, and $\mathbf x_{i,t}^{\mathrm{int}}$ each group the dynamic and static traffic features listed in Table~\ref{tab:upper-traffic-features}.
The remaining terms describe the candidate plans: $\mathbf X_{i,t}^{\mathrm{cand}}$ contains candidate features, while $\mathbf U_{i,t}^{\mathrm{stage}}$, $\mathbf M_{i,t}^{\mathrm{stage}}$, and $\boldsymbol\mu_{i,t}$ represent their stage sequences and validity masks, as detailed in Table~\ref{tab:upper-candidate-features}.
Dynamic tensors are padded to at most $\bar M$ lanes, $\bar P$ phases, $K$ candidates, and $\bar R$ stages. The experiments use $(\bar M,\bar P,K,\bar R)=(24,8,25,8)$. The lane--phase graph is cached for each network.

\paragraph{Traffic representation}
Table~\ref{tab:upper-traffic-features} summarizes the lane-, phase-, and intersection-level features. The point and spatial queues used as policy features are derived from $x_i(k_t)$ through Equations~\eqref{eq:point-queue-dynamics} and~\eqref{eq:spatial-queue-dynamics}. Let $C_{i,m}$ denote the storage capacity of lane $m$, let $s_{i,m}$ be its saturation discharge capacity per micro step as defined in Section~\ref{sec:lower-queue-models}, and let $B^{\mathrm{arr}}$ be the number of arrival bins covering the prediction horizon. For arrival bin $\beta$ with prediction-offset set $\mathcal K_\beta^{\mathrm{arr}}$, define $\hat a_{i,m,\beta}=\sum_{p\in\mathcal K_\beta^{\mathrm{arr}}}\hat q_{i,m}^{\mathrm{arr}}(p\mid k_t)$ and $\hat a_{i,m}=\sum_{p=0}^{H-1}\hat q_{i,m}^{\mathrm{arr}}(p\mid k_t)$. The active phase indicator is $\chi_{i,t}(\phi)=\mathbf 1\{\phi=\phi_i^{\mathrm{act}}\}$, and $\vartheta_{i,t}(\phi)$ is the cyclic angular displacement of phase $\phi$ from $\phi_i^{\mathrm{act}}$. Static attributes are cached with $\mathcal G_i^{\mathrm{lp}}$, while invalid padded rows are excluded by the lane and phase masks.

\begin{table}
\centering
\setlength{\tabcolsep}{4pt}
\renewcommand{\arraystretch}{1.0}
\caption{Traffic state inputs to the candidate selector.}
\label{tab:upper-traffic-features}
\begin{tabular}{@{}
c
c
>{\raggedright\arraybackslash}p{0.39\linewidth}
>{\raggedright\arraybackslash}p{0.28\linewidth}@{}}
\toprule
Symbol & Shape & Description & Normalization \\
\midrule
 $\mathbf X_{i,t}^{\mathrm{lane,dyn}}$
 & $\bar M\times(B^{\mathrm{arr}}+3)$
 & \textbf{Lane dynamic:} Point and spatial queues; $B^{\mathrm{arr}}$ binned predicted arrival counts; predicted arrival count over the horizon
 & $Q^{\mathrm p}/C$, $Q^{\mathrm s}/C$, $\hat a_\beta/(s|\mathcal K_\beta^{\mathrm{arr}}|)$, and $\hat a/(sH)$ \\
\addlinespace[2pt]

$\mathbf X_i^{\mathrm{lane,stat}}$
& $\bar M\times6$
& \textbf{Lane static:} Effective link length and a five-dimensional multi-hot encoding of lane movements
& Length divided by $300$~m; movement indicators are binary \\
\addlinespace[2pt]

$\mathbf X_{i,t}^{\mathrm{phase,dyn}}$
& $\bar P\times6$
& \textbf{Phase dynamic:} Active phase indicator, elapsed green, remaining minimum and maximum green, and relative phase sine/cosine
& Indicator is binary; green times divided by $g^{\max}$ or $H$; sine/cosine lie in $[-1,1]$ \\
\addlinespace[2pt]

$\mathbf X_i^{\mathrm{phase,stat}}$
& $\bar P\times2$
& \textbf{Phase static:} Minimum and maximum green times
& Both divided by $H$ \\
\addlinespace[2pt]

$\mathbf x_{i,t}^{\mathrm{int,dyn}}$
& $5$
& \textbf{Intersection dynamic:} Intergreen indicator and remaining duration; aggregate point queue, spatial queue, and predicted arrival count over the horizon
& Indicator is binary; duration divided by $H$; queues divided by aggregate storage capacity; arrivals divided by aggregate saturation discharge over the horizon \\
\addlinespace[2pt]

$\mathbf x_i^{\mathrm{int,stat}}$
& $2$
& \textbf{Intersection static:} Fractions of valid lane and phase slots
& $|\mathcal M_i|/\bar M$ and $|\Phi_i|/\bar P$ \\
\bottomrule
\end{tabular}
\end{table}

Lane features retain lane-specific queue and arrival information, while phase features describe the current signal timing. The lane--phase graph relates these features through the service relation, and intersection features provide aggregate traffic conditions. The normalizations in Table~\ref{tab:upper-traffic-features} keep feature magnitudes comparable across intersections.

\paragraph{Candidate plan representation}
Each candidate is described by its predicted objective trade-off and its ordered timing structure. Let
$\mathbf F_i^{(c)}(k_t)=[F_i^{\mathrm{delay},(c)}(k_t),F_i^{\mathrm{queue},(c)}(k_t),F_i^{\mathrm{stop},(c)}(k_t)]^\top$
contain the predicted values of total queueing delay, peak queue accumulation, and total number of stops defined in Section~\ref{sec:lower-objective-computation}. For objective component $\rho\in\{\mathrm{delay},\mathrm{queue},\mathrm{stop}\}$, the absolute objective is transformed by $\log(1+F_{i,\rho}^{(c)}(k_t))$. Its within-set relative score is
\begin{equation}
 f_{i,\mathrm{rel},\rho}^{(c)}(k_t)
=
\begin{cases}
\left(
1-2\dfrac{F_{i,\rho}^{(c)}(k_t)-F_{i,\rho}^{\min}(k_t)}
{F_{i,\rho}^{\max}(k_t)-F_{i,\rho}^{\min}(k_t)}
\right),
& F_{i,\rho}^{\max}(k_t)>F_{i,\rho}^{\min}(k_t),
\\[8pt]
0,
& F_{i,\rho}^{\max}(k_t)=F_{i,\rho}^{\min}(k_t),
\end{cases}
\label{eq:upper-relative-objective-feature}
\end{equation}
where $F_{i,\rho}^{\min}(k_t)$ and $F_{i,\rho}^{\max}(k_t)$ are respectively the minimum and maximum of objective component $\rho$ over the $K'_i(t)$ non-padded candidates. Thus, the best and worst values of an objective map to $1$ and $-1$, respectively, while an objective with no within-set variation contributes zero.

The candidate feature tensor $\mathbf X_{i,t}^{\mathrm{cand}}$ combines these absolute and relative objective features with the number of phase changes. The stage tensor $\mathbf U_{i,t}^{\mathrm{stage}}$ retains the ordered timing structure of each plan. Its seven scalar features are green duration normalized by the phase-specific maximum green, green duration normalized by $H$, stage start time normalized by $H$, normalized deviation from the corresponding horizon-shifted reference phase end time, a reference-boundary indicator, and the sine and cosine of the phase position relative to the active phase. Table~\ref{tab:upper-candidate-features} lists these tensors together with the masks used to distinguish padding from genuine zero-valued features.

\begin{table}
\centering
\setlength{\tabcolsep}{4pt}
\renewcommand{\arraystretch}{1.0}
\caption{Candidate plan features and masks.}
\label{tab:upper-candidate-features}
\begin{tabular}{@{}
c
c
>{\raggedright\arraybackslash}p{0.40\linewidth}
>{\raggedright\arraybackslash}p{0.29\linewidth}@{}}
\toprule
Symbol & Shape & Description & Normalization \\
\midrule
$\mathbf X_{i,t}^{\mathrm{cand}}$
& $K\times7$
& \textbf{Candidate features:} Three absolute objectives, three within-set relative scores, and the number of phase changes
 & Absolute objectives use $\log(1+F)$; relative scores lie in $[-1,1]$; phase changes are divided by $\max(\bar R-1,1)$ \\
\addlinespace[2pt]

$\mathbf U_{i,t}^{\mathrm{stage}}$
& $K\times\bar R\times7$
 & \textbf{Stage features:} Green duration divided by $g^{\max}$; green duration divided by $H$; normalized stage start; reference boundary deviation; reference indicator; relative phase sine and cosine
 & The two duration features use $g^{\max}$ and $H$; start time is divided by $H$; deviation is divided by the allowed boundary shift; the indicator is binary; sine/cosine lie in $[-1,1]$ \\
\addlinespace[2pt]

$\mathbf M_{i,t}^{\mathrm{stage}}$
& $K\times\bar R$
& \textbf{Stage mask:} Identifies non-padded stages
& Binary \\
\addlinespace[2pt]

$\boldsymbol\mu_{i,t}$
& $K$
& \textbf{Candidate mask:} Identifies the $K'_i(t)$ non-padded candidate slots
& Binary \\
\bottomrule
\end{tabular}
\end{table}

Candidate features summarize the predicted objectives and phase-change count of each candidate. Stage features retain the candidate's ordered phase-timing sequence, allowing candidates with similar objective vectors to remain distinguishable. Zero padding and the stage mask identify valid stages within each candidate, while the candidate mask zeros padded candidate embeddings and subsequently excludes those slots from the actor and critic.

\subsection{Observation encoding}
\label{sec:upper-topology-encoder}

\begin{figure}
\centering
\begin{subfigure}[t]{\linewidth}
\centering
\includegraphics[width=0.6\linewidth]{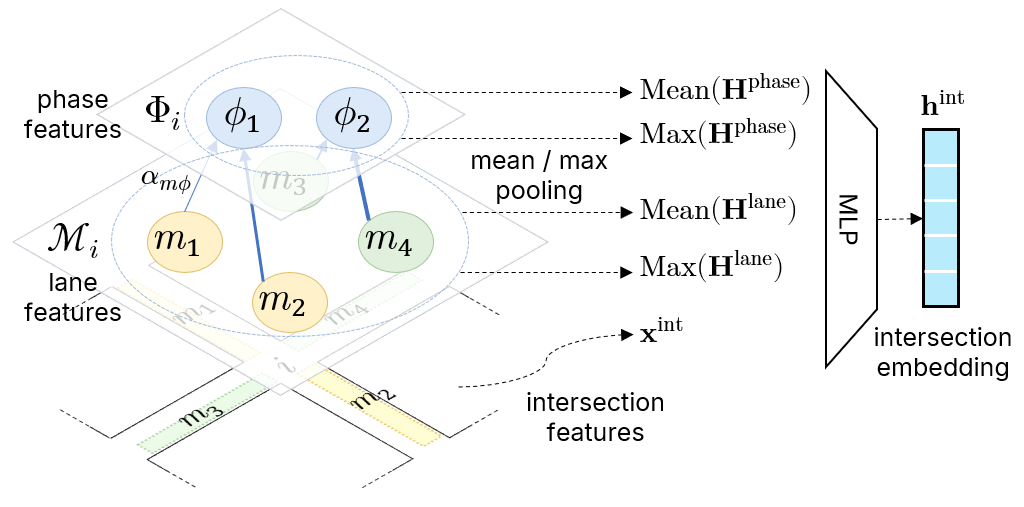}
\caption{Topology-aware intersection encoder.}
\label{fig:upper-intersection-encoder}
\end{subfigure}

\vspace{4pt}

\begin{subfigure}[t]{\linewidth}
\centering
\includegraphics[width=0.6\linewidth]{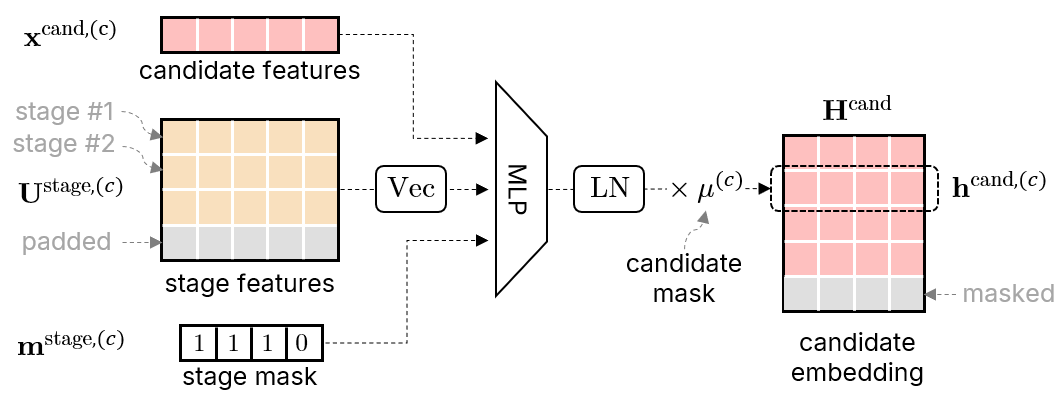}
\caption{Shared per-candidate encoder.}
\label{fig:upper-candidate-encoder}
\end{subfigure}
\caption{Observation encoders used by the candidate selector. The intersection encoder aggregates lane embeddings along the lane--phase graph and pools the valid lane and phase embeddings with intersection features to obtain $\mathbf h_{i,t}^{\mathrm{int}}$. The candidate encoder combines candidate-level features, zero-padded stage features, and the stage mask, then zeros padded candidate embeddings with the candidate mask to produce $\mathbf H_{i,t}^{\mathrm{cand}}$. The actor and critic instantiate the same architecture with separate parameters and reuse the candidate mask in their downstream scoring and aggregation operations.}
\label{fig:upper-observation-encoders}
\end{figure}

The actor and critic instantiate separate observation encoders with the same structure, as illustrated in Fig.~\ref{fig:upper-observation-encoders}. Each encoder contains a topology-aware intersection branch and a shared per-candidate branch. To avoid duplicating the encoding equations, the network superscripts $\pi$ and $V$ are suppressed throughout this subsection. Within each network, multilayer perceptron (MLP) parameters are shared across agents. The traffic encoder first maps each valid lane and phase to $\mathbf h_{i,m,t}^{\mathrm{lane}}$ and $\mathbf h_{i,\phi,t}^{\mathrm{phase},0}$, respectively. A lane-to-phase graph attention layer then aggregates only physically admissible service relations. For $(m,\phi)\in\mathcal E_i^{\mathrm{lp}}$,
\begin{equation}
\begin{aligned}
e_{i,m\phi,t}
&=
\mathbf w_{\mathrm{att}}^\top
\operatorname{LReLU}\!\left(
\mathbf W_{\mathrm l}\mathbf h_{i,m,t}^{\mathrm{lane}}
+
\mathbf W_{\mathrm p}\mathbf h_{i,\phi,t}^{\mathrm{phase},0}
\right),
\\
\alpha_{i,m\phi,t}
&=
\operatorname{softmax}_{m\in\mathcal M_i(\phi)}(e_{i,m\phi,t}),
\qquad
\mathbf g_{i,\phi,t}
=
\sum_{m\in\mathcal M_i(\phi)}
\alpha_{i,m\phi,t}
\mathbf W_{\mathrm l}\mathbf h_{i,m,t}^{\mathrm{lane}}.
\end{aligned}
\label{eq:upper-lane-phase-attention}
\end{equation}
The learned summary $\mathbf g_{i,\phi,t}$ is complemented by $\mathbf z_{i,\phi,t}^{\mathrm{agg}}$, which contains phase-level normalized aggregates of the point queue, spatial queue, and predicted arrivals over the lanes served by phase $\phi$, together with the fraction of local lanes served. Let $\mathbf H_{i,t}^{\mathrm{lane}}$ collect the valid lane embeddings and let $\mathbf H_{i,t}^{\mathrm{phase}}$ collect the updated phase embeddings. The phase and intersection updates are
\begin{equation}
\begin{aligned}
\mathbf h_{i,\phi,t}^{\mathrm{phase}}
&=
\operatorname{LN}\!\left(
\mathbf h_{i,\phi,t}^{\mathrm{phase},0}
+f_{\mathrm{upd}}([
\mathbf g_{i,\phi,t},
\mathbf z_{i,\phi,t}^{\mathrm{agg}}
])
\right),
\\
\mathbf h_{i,t}^{\mathrm{int}}
&=
f_{\mathrm{int}}\!\left([
\operatorname{Mean}(\mathbf H_{i,t}^{\mathrm{lane}}),
\operatorname{Max}(\mathbf H_{i,t}^{\mathrm{lane}}),
\operatorname{Mean}(\mathbf H_{i,t}^{\mathrm{phase}}),
\operatorname{Max}(\mathbf H_{i,t}^{\mathrm{phase}}),
\mathbf x_{i,t}^{\mathrm{int}}
]
\right).
\end{aligned}
\label{eq:upper-intersection-embedding}
\end{equation}
Mean pooling controls for intersection size, whereas maximum pooling retains locally dominant lane or phase conditions.

In parallel, the candidate encoder concatenates candidate features, the flattened zero-padded stage features, and the stage mask. Let $\mathbf x_{i,t}^{\mathrm{cand},(c)}$ denote row $c$ of $\mathbf X_{i,t}^{\mathrm{cand}}$, and let $\mathbf U_{i,t}^{\mathrm{stage},(c)}$ and $\mathbf m_{i,t}^{\mathrm{stage},(c)}$ denote the corresponding candidate slices of $\mathbf U_{i,t}^{\mathrm{stage}}$ and $\mathbf M_{i,t}^{\mathrm{stage}}$, respectively:
\begin{equation}
\mathbf h_{i,t}^{\mathrm{cand},(c)}
=
\mu_{i,t}^{(c)}
\operatorname{LN}\!\left(
f_{\mathrm{cand}}\!\left([
\mathbf x_{i,t}^{\mathrm{cand},(c)},
\operatorname{vec}\!\left(\mathbf U_{i,t}^{\mathrm{stage},(c)}\right),
\mathbf m_{i,t}^{\mathrm{stage},(c)}
]
\right)
\right).
\label{eq:upper-candidate-embedding}
\end{equation}
Here, $\operatorname{vec}(\cdot)$ stacks the zero-padded stage-feature matrix into a vector, while the stage mask distinguishes valid stages from padding. The prefactor $\mu_{i,t}^{(c)}$ sets the embedding of a padded candidate slot to zero. Applying the same mapping to all $K$ candidate slots yields the candidate-embedding matrix $\mathbf H_{i,t}^{\mathrm{cand}}$. The actor and critic subsequently reuse the candidate mask to exclude padded slots from candidate scoring and aggregation.

\subsection{Actor and critic}
\label{sec:upper-actor-critic}

The actor uses a separate observation encoder and attention head from the critic, with each network shared across agents. It scores each candidate against the intersection embedding. Let $\mathbf q_{i,t}^{\pi}=\mathbf W_{\mathrm q}^{\pi}\mathbf h_{i,t}^{\mathrm{int}}$ and $\mathbf k_{i,t}^{\pi,(c)}=\mathbf W_{\mathrm k}^{\pi}\mathbf h_{i,t}^{\mathrm{cand},(c)}$. The masked logit and categorical policy are
\begin{equation}
z_{i,t}^{\pi,(c)}
=
\begin{cases}
(\mathbf q_{i,t}^{\pi})^\top\mathbf k_{i,t}^{\pi,(c)}/\sqrt{d_{\mathrm{att}}},
& \mu_{i,t}^{(c)}=1,
\\
-\infty,
& \mu_{i,t}^{(c)}=0,
\end{cases}
\qquad
\pi_{\theta_\pi}(a_{i,t}=c\mid o_{i,t})
=
\frac{\exp(z_{i,t}^{\pi,(c)})}
{\sum_{c'=1}^{K}\exp(z_{i,t}^{\pi,(c')})}.
\label{eq:actor-policy}
\end{equation}
Here, $d_{\mathrm{att}}$ is the query and key dimension. Applying the same candidate encoder to every candidate slot makes the policy equivariant to candidate permutations, and the candidate mask serves as an action mask that assigns exactly zero probability to padded candidates.

The critic applies the same masked scaled dot-product operation as the actor, using separate query and key projections. Let $z_{i,t}^{V,(c)}$ denote the resulting scores and $\alpha_{i,t}^{V,(c)}$ their softmax weights over non-padded candidates. Candidate values are projected as $\mathbf v_{i,t}^{V,(c)}=\mathbf W_{\mathrm v}^{V}\mathbf h_{i,t}^{\mathrm{cand},(c)}$. The local value is computed from the intersection embedding and four summaries of the non-padded candidates:
\begin{equation}
\begin{aligned}
\mathbf h_{i,t}^{\mathrm{ctx}}
&=
\sum_{c=1}^{K}
\alpha_{i,t}^{V,(c)}
\mathbf v_{i,t}^{V,(c)},
\qquad
\bar{\mathbf h}_{i,t}^{\mathrm{cand}}
=
\frac{1}{K'_i(t)}
\sum_{c=1}^{K}
\mu_{i,t}^{(c)}\mathbf h_{i,t}^{\mathrm{cand},(c)},
\\
V_{\theta_V}(o_{i,t})
&=
f_V\!\left([
\mathbf h_{i,t}^{\mathrm{int}},
\mathbf h_{i,t}^{\mathrm{ctx}},
\bar{\mathbf h}_{i,t}^{\mathrm{cand}},
\max_{c:\mu_{i,t}^{(c)}=1}z_{i,t}^{V,(c)},
K'_i(t)/K
]
\right).
\end{aligned}
\label{eq:critic-value}
\end{equation}
Here, $\mathbf h_{i,t}^{\mathrm{ctx}}$ is the attention-weighted candidate embedding, and $\bar{\mathbf h}_{i,t}^{\mathrm{cand}}$ is the mean embedding of the non-padded candidates. The maximum score records the strongest match between the traffic context and a candidate plan, while $K'_i(t)/K$ gives the fraction of non-padded candidates. These inputs are unchanged by candidate permutations, so the value estimate is invariant to candidate indexing.

\subsection{Reward}
\label{sec:upper-reward-design}

The reward used to train the candidate selector's policy evaluates realized rather than model-predicted performance. Let $Q_{i,m}^{\mathrm{obs}}(k)$ be the stopped vehicle count observed on inbound lane $m$ at micro step $k$, and let $\mathcal K_t^{\mathrm{exec}}=\{k_t+1,\dots,k_t+M\}$ be the executed interval. The local reward is
\begin{equation}
r_{i,t}
=
-\Delta_k
\sum_{k\in\mathcal K_t^{\mathrm{exec}}}
\sum_{m\in\mathcal M_i}
Q_{i,m}^{\mathrm{obs}}(k).
\label{eq:upper-reward}
\end{equation}
It is measured in vehicle-time units; because $\Delta_k=1$~s in the experiments, the implemented sum is numerically equal to vehicle-seconds. The observed queue is distinguished from the model-propagated point queue $Q_{i,m}^{\mathrm p}$ used by SP-MODP.

The two control levels use different evaluation horizons. SP-MODP evaluates total queueing delay, peak queue accumulation, and total number of stops over the full prediction horizon to generate a diverse feasible set, whereas the reward measures realized queueing delay only over the executed interval. Prediction errors, traffic interactions, and future replanning can therefore make the candidate plan with the smallest predicted queueing delay differ from the one that maximizes long-term return. The reward therefore trains the candidate selector's policy to choose among feasible candidate plans under repeated receding horizon execution.

\subsection{Training and execution}
\label{sec:upper-training-deployment}

We train the selector using IPPO with parameter sharing~\citep{schulman2017proximal,dewitt2020independent,yu2022surprising}. Each intersection uses only its local observation and local value estimate; neither network receives a centralized state. Rollouts from all agents and parallel simulation environments are pooled to update one actor and one critic. Advantages are estimated with generalized advantage estimation (GAE)~\citep{schulman2016highdimensional}. For each agent, the advantages are standardized to zero mean and unit variance across rollout workers and time steps.

The actor is optimized with the clipped PPO surrogate and an entropy bonus. The critic uses the Huber loss~\citep{huber1964robust} on PopArt-normalized value targets~\citep{hasselt2016learning}. The final value layer is rescaled as the running target statistics change so that predictions in the original return scale are preserved.
Separate actor and critic optimizers, gradient clipping, and approximate KL early stopping are used. Detailed hyperparameters are reported in \ref{app:training-details}.

During training, actions are sampled from Equation~\eqref{eq:actor-policy}; during evaluation and deployment, the feasible candidate plan with the largest probability is selected. Every selectable action satisfies the phase order, green time, intergreen, and inter-update timing constraints of the candidate generation problem.

\section{Experiments}
\label{sec:experiments}

\subsection{Experimental setup}
\label{sec:experimental-setup}

The experiments are designed to address four research questions:

\begin{enumerate}
\item[\textbf{RQ1}] How does SelectLight compare with model-based, direct RL, and traffic-responsive baselines, and does its advantage persist as demand increases?
\item[\textbf{RQ2}] How well does the learned policy transfer without fine-tuning across demand levels and networks?
\item[\textbf{RQ3}] What component-level and operational evidence explains SelectLight's performance gains?
\item[\textbf{RQ4}] Can SP-MODP generate candidate plans within the online control budget?
\end{enumerate}

\subsubsection{Scenarios}

\begin{figure}
\centering
\begin{subfigure}[t]{0.56\textwidth}
\centering
\includegraphics[width=\linewidth]{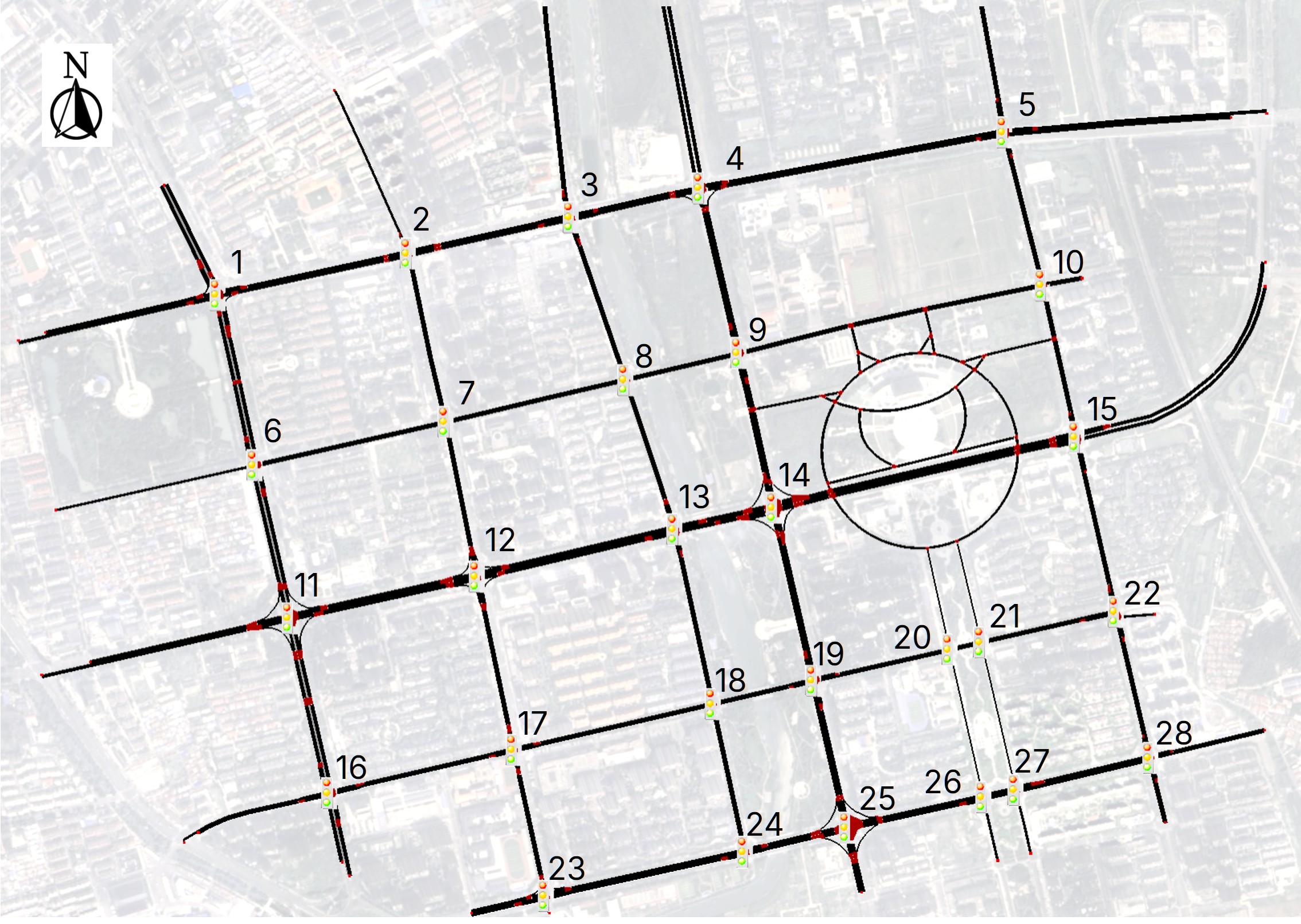}
\caption{Lianyungang network.}
\label{fig:network-lianyungang}
\end{subfigure}
\hfill
\begin{subfigure}[t]{0.38\textwidth}
\centering
\includegraphics[width=\linewidth]{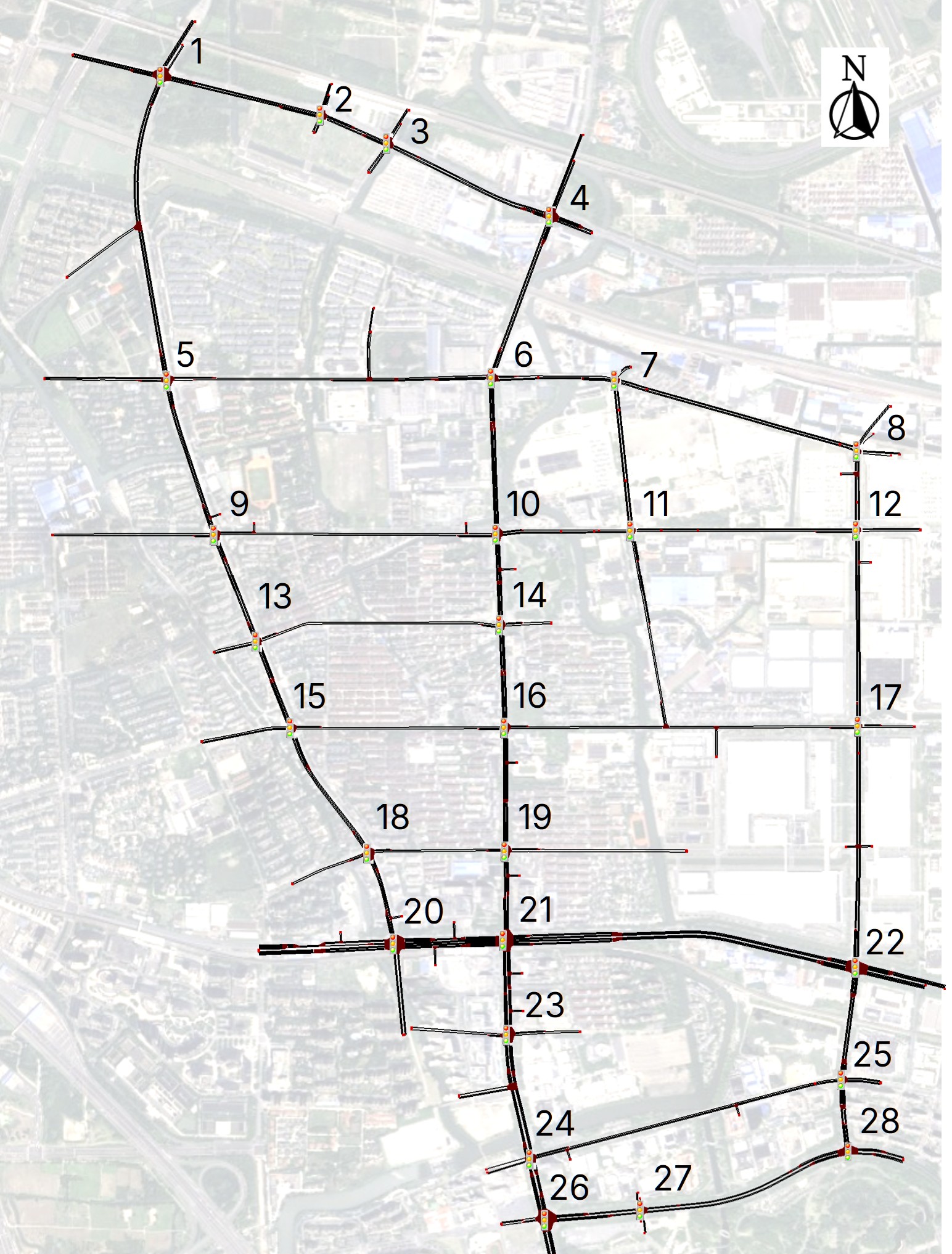}
\caption{Jiading network.}
\label{fig:network-jiading}
\end{subfigure}
\caption{SUMO networks representing urban areas in (a) Lianyungang, Jiangsu Province, and (b) Jiading District, Shanghai. Numbers identify the controlled signalized intersections.}
\label{fig:experimental-networks}
\end{figure}

We evaluate the controllers on two 28-intersection SUMO networks modeled on real urban areas in China: Lianyungang, Jiangsu Province, and Jiading District, Shanghai (Fig.~\ref{fig:experimental-networks}). Both networks retain heterogeneous intersection geometries and native phase sequences. They include channelized right-turn lanes and shared-use lanes, with different numbers of phases and protected-movement combinations. Together, these features capture signal control complexity that is often simplified in idealized grid networks.

The Lianyungang network is the primary setting for evaluating demand robustness. Its nominal input increases from 6,350~veh/h at scale 1.0 to 9,525 and 12,700~veh/h at scales 1.5 and 2.0. The Jiading network is evaluated at scale 1.0 as a complementary setting with a different topology and demand pattern. Because both networks contain 28 controlled intersections, the comparison focuses on differences in network structure and demand distribution, with network size held constant.

Traffic demands are represented as fixed path-flow inputs, consistent with simulation-based evaluations that derive route demands through origin--destination or path-flow estimation \citep{chen2021dynamic,tang2021dynamic}. Table~\ref{tab:network-demand-characteristics} characterizes both networks at the reference scale of 1.0. Intersection capacity utilization (ICU) is computed using the critical-movement procedure in \citet{husch2003intersection} and \citet{dowling2007traffic}. Lianyungang has the higher network input flow, whereas Jiading has the higher mean and median intersection load. The maximum ICU corresponds to level-of-service B in both networks. These complementary scenarios support tests of performance under increasing demand and zero-shot transfer to a different urban network.

\begin{table}
\centering
\caption{Demand and intersection-load characteristics of the experimental networks at demand scale 1.0.}
\label{tab:network-demand-characteristics}
\begin{threeparttable}
\begin{tabular}{lcccccc}
\toprule
\multirow{2}{*}{Network} &
\multirow{2}{*}{Input flow (veh/h)} &
\multicolumn{4}{c}{ICU (\%)} &
\multirow{2}{*}{LOS at maximum ICU} \\
\cmidrule(lr){3-6}
& & Min. & Mean & Median & Max. & \\
\midrule
Lianyungang & 6,350 & 30.8 & 43.4 & 46.7 & 63.3 & B \\
Jiading & 4,839 & 39.2 & 52.5 & 54.4 & 63.3 & B \\
\bottomrule
\end{tabular}
\begin{tablenotes}
\footnotesize
\item ICU = intersection capacity utilization; LOS = level of service.
\end{tablenotes}
\end{threeparttable}
\end{table}

\subsubsection{Compared methods}

We compare SelectLight with six baselines spanning model-based, direct RL, and traffic-responsive control:

\begin{itemize}
\item \textbf{SelectLight} uses SP-MODP to generate a bounded set of mutually nondominated signal plans and an IPPO-trained policy to select one plan.

\item \textbf{Multi-objective DMPC (MO-DMPC)} uses the same prediction module and candidate generators as SelectLight but replaces the learned selector with a fixed minimum-distance rule in normalized objective space. This comparison holds the prediction and candidate-generation mechanisms fixed while changing the selection rule.

\item \textbf{Single-objective DMPC (SO-DMPC)} uses the same prediction module and signal timing constraints but directly optimizes predicted total queueing delay. It does not perform multi-objective candidate generation or learned candidate selection.

\item \textbf{IPPO} directly chooses whether to retain the current phase or switch to the next admissible phase using current detector observations.

\item \textbf{GAT-IPPO}, inspired by CoLight \citep{wei2019colight}, augments IPPO with graph attention among directly connected intersections while retaining the same local representation and phase-switching action.

\item \textbf{Actuated} extends or terminates the current phase in response to detector measurements.

\item \textbf{MP} selects the admissible phase with the largest queue pressure \citep{varaiya2013max}.
\end{itemize}

IPPO and GAT-IPPO test direct phase switching against SelectLight's structured candidate-plan action space. MO-DMPC compares learned and fixed selection rules under a common prediction and candidate-generation design, whereas SO-DMPC tests the combined value of multi-objective candidate generation and learned selection over a delay-aligned predictive controller. Further implementation details and a structural comparison appear in \ref{app:learning-method-implementation} and Table~\ref{tab:learning-method-structures}, respectively.

\subsubsection{Evaluation metrics}

We report one intersection-level metric and four trip-level metrics:

\begin{itemize}
\item \textbf{Average cumulative queue (ACQ)} is the cumulative observed point queue over an episode, averaged across controlled intersections:
\begin{equation}
\mathrm{ACQ}
=
\frac{1}{|\mathcal V|}
\sum_{i\in\mathcal V}
\sum_{k=1}^{K_{\mathrm{ep}}}
\sum_{m\in\mathcal M_i}
Q_{i,m}^{\mathrm{obs}}(k),
\label{eq:evaluation-acq}
\end{equation}
where $K_{\mathrm{ep}}$ is the number of simulation steps. At the 1~s resolution used here, ACQ equals realized queueing delay in vehicle-seconds per controlled intersection and is directly aligned with the upper-level reward.

\item \textbf{Average travel time (ATT)} is the mean trip time of vehicles that complete their journeys during the evaluation episode.

\item \textbf{Average waiting time (AWT)} is the mean accumulated waiting time of completed vehicles.

\item \textbf{Average stop count (ASC)} is the mean number of transitions from motion to a waiting state among completed vehicles.

\item \textbf{Throughput (THP)} is the number of completed trips. It complements the completed-vehicle averages by revealing whether a controller leaves more trips unfinished at the end of the episode.
\end{itemize}

\subsubsection{Implementation and evaluation protocol}

All experiments were conducted in the SUMO microscopic traffic simulator \citep{lopez2018microscopica} through \texttt{libsumo}. The learning methods were implemented with \texttt{PyTorch}, and the SP-MODP solver was implemented in C++. Experiments were run under WSL2 on a workstation with an Intel Core i9-13900K CPU, an NVIDIA GeForce RTX 4090 GPU, and 64~GB RAM. Table~\ref{tab:control-settings} summarizes the common temporal settings and the settings specific to candidate generation. The prediction module was calibrated before policy training and then held fixed. Evaluations use the final checkpoint from each training run; no checkpoint selection, fine-tuning, or test-time adaptation is applied.

\begin{table}
\centering
\caption{Temporal, signal control, and candidate-generation settings.}
\label{tab:control-settings}
\begin{tabular}{ll}
\toprule
Parameter & Value \\
\midrule
Simulation step & 1~s \\
Macro decision interval $M$ & 5~s \\
Prediction horizon $H$ & 120~s \\
Episode length & 3600~s \\
Minimum green & 10~s \\
Maximum green & 80~s \\
Intergreen & 3~s \\
Maximum inter-update phase-end adjustment & 10~s \\
SP-MODP timing discretization & 2~s \\
Maximum stages $\bar R$ & 8 \\
Maximum candidate set size $K$ & 25 \\
Per-node label cap $K_i^{\mathrm{pf}}$ & 50 \\
\bottomrule
\end{tabular}
\end{table}

SelectLight, MO-DMPC, SO-DMPC, IPPO, and GAT-IPPO update their controls every 5~s, whereas MP updates every 15~s. Actuated control responds at the 1~s simulation resolution with a 3~s maximum gap. All methods use minimum and maximum green times of 10 and 80~s, retain the native phase sequences of both networks, and use a 3~s intergreen.

Unless stated otherwise, the performance tables in this section and \ref{app:additional-experimental-results} report the mean $\pm$ sample standard deviation over eight evaluation seeds. Bold and underlined values denote the best and second-best result within each comparison group, respectively; arrows indicate the preferred direction. For demand robustness and zero-shot transfer, relative changes are computed between runs with the same seed and then averaged. Training curves show one final run per method and network.

\subsection{Overall performance}
\label{sec:overall-performance}

To address RQ1, we first compare SelectLight with all baselines on both experimental networks.

\begin{table}
\centering
\caption{Overall performance on the two experimental networks.}
\label{tab:overall-performance}
\resizebox{\textwidth}{!}{%
\begin{tabular}{llccccc}
\toprule
Network & Method & ACQ $\downarrow$ & ATT $\downarrow$ & AWT $\downarrow$ & ASC $\downarrow$ & THP $\uparrow$ \\
\midrule
\multirow{7}{*}{Lianyungang} & Actuated & $12532.57 \pm 201.77$ & $300.22 \pm 1.41$ & $54.48 \pm 0.86$ & $3.202 \pm 0.042$ & $5827.25 \pm 18.43$ \\
 & MP & $16192.09 \pm 227.21$ & $314.02 \pm 2.38$ & $70.37 \pm 1.00$ & $3.072 \pm 0.025$ & $5786.50 \pm 12.73$ \\
 & MO-DMPC & $11631.25 \pm 295.06$ & $292.84 \pm 1.93$ & $50.66 \pm 1.35$ & \underline{$2.545 \pm 0.014$} & $5837.38 \pm 15.49$ \\
 & SO-DMPC & \underline{$9886.19 \pm 159.47$} & \underline{$284.61 \pm 1.67$} & \underline{$42.86 \pm 0.69$} & $2.570 \pm 0.030$ & \underline{$5849.12 \pm 9.86$} \\
 & IPPO & $11683.03 \pm 248.63$ & $293.31 \pm 2.26$ & $50.74 \pm 1.06$ & $2.734 \pm 0.038$ & $5830.62 \pm 12.34$ \\
 & GAT-IPPO & $11114.40 \pm 179.16$ & $290.25 \pm 2.32$ & $48.15 \pm 0.79$ & $2.695 \pm 0.047$ & $5840.50 \pm 11.51$ \\
 & SelectLight & {\bfseries\boldmath $9720.03 \pm 77.03$} & {\bfseries\boldmath $283.38 \pm 1.47$} & {\bfseries\boldmath $42.10 \pm 0.35$} & {\bfseries\boldmath $2.494 \pm 0.017$} & {\bfseries\boldmath $5851.62 \pm 14.28$} \\
\midrule
\multirow{7}{*}{Jiading} & Actuated & $12507.93 \pm 330.33$ & $340.66 \pm 1.98$ & $70.42 \pm 1.82$ & $3.803 \pm 0.069$ & $4383.38 \pm 7.73$ \\
 & MP & $15802.15 \pm 232.32$ & $354.11 \pm 1.29$ & $89.26 \pm 1.24$ & $3.250 \pm 0.021$ & $4357.38 \pm 7.60$ \\
 & MO-DMPC & $12961.55 \pm 117.16$ & $338.08 \pm 1.07$ & $74.10 \pm 0.80$ & {\bfseries\boldmath $2.961 \pm 0.027$} & $4386.62 \pm 7.87$ \\
 & SO-DMPC & \underline{$10456.39 \pm 134.55$} & \underline{$324.74 \pm 1.21$} & \underline{$59.11 \pm 0.89$} & $3.101 \pm 0.031$ & {\bfseries\boldmath $4403.50 \pm 7.54$} \\
 & IPPO & $11357.40 \pm 165.36$ & $330.61 \pm 1.24$ & $64.12 \pm 1.03$ & $3.193 \pm 0.023$ & $4390.00 \pm 5.71$ \\
 & GAT-IPPO & $10960.83 \pm 198.82$ & $328.61 \pm 1.51$ & $61.81 \pm 1.29$ & $3.190 \pm 0.025$ & $4397.75 \pm 6.86$ \\
 & SelectLight & {\bfseries\boldmath $10256.01 \pm 156.08$} & {\bfseries\boldmath $323.62 \pm 1.26$} & {\bfseries\boldmath $57.96 \pm 0.93$} & \underline{$3.062 \pm 0.019$} & \underline{$4401.25 \pm 6.32$} \\
\bottomrule
\end{tabular}}
\end{table}

Overall, SelectLight outperforms the baselines on most metrics across both networks (Table~\ref{tab:overall-performance}). It ranks first on all five metrics in Lianyungang and on ACQ, ATT, and AWT in Jiading, while ranking second on ASC and THP there.

Among the baselines, SO-DMPC achieves the best ACQ, ATT, and AWT on both networks, making it the most relevant reference for assessing SelectLight's delay-related gains. SelectLight reduces ACQ and AWT by 1.68--1.95\% relative to SO-DMPC across the two networks, while its ATT reduction remains below 0.5\%. These modest margins show that direct delay optimization is already highly effective, with SelectLight providing a consistent additional gain.

Comparing both DMPC baselines further clarifies this advantage. SO-DMPC closes most of the large gap between MO-DMPC and SelectLight, yet SelectLight remains best on ACQ, ATT, and AWT in both networks. SelectLight therefore preserves the strength of direct delay optimization while achieving the strongest overall balance across the full metric set.

Among the remaining baselines, GAT-IPPO consistently improves upon IPPO on ACQ, ATT, and AWT on both networks, showing the value of exchanging information across adjacent intersections for direct phase-switching policies. Both learning baselines also outperform Actuated and MP on the three delay-related metrics. MP records the largest delays on both networks, indicating that local queue-pressure maximization is less effective than learned or predictive coordination for the tested network-wide objective.

\subsection{Robustness across demand levels}
\label{sec:demand-robustness}

\begin{figure}
\centering
\includegraphics[width=\textwidth]{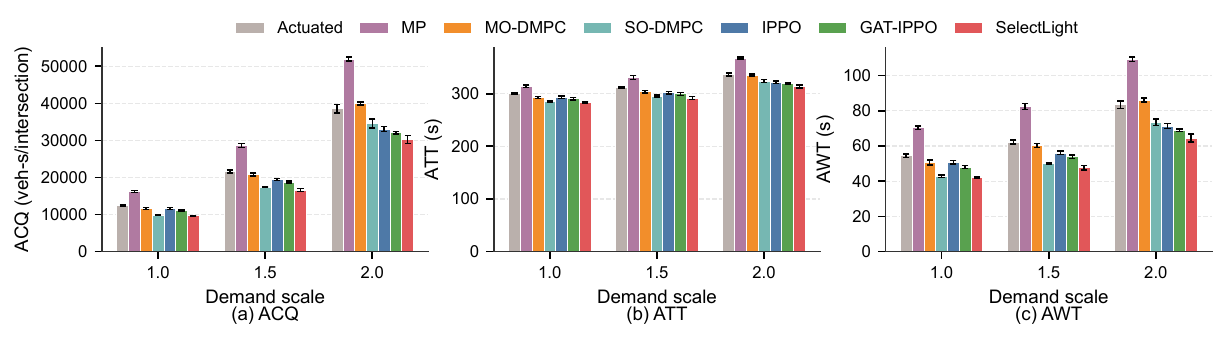}
\caption{Performance across demand levels on the Lianyungang network. Each learning method is trained at the corresponding demand scale; non-learning methods require no training. Error bars denote sample standard deviations over $N=8$ evaluation seeds.}
\label{fig:demand-robustness}
\end{figure}

To further address RQ1, we evaluate how controller performance changes with traffic load at demand scales 1.0, 1.5, and 2.0 on Lianyungang. Each learning method is trained and evaluated at every scale. Figure~\ref{fig:demand-robustness} reports the delay-related metrics, and Table~\ref{tab:app-demand-robustness} provides all five metrics.

Increasing demand separates the controllers more clearly and changes which baseline is most competitive. SO-DMPC is the strongest baseline at scales 1.0 and 1.5, but its ACQ gap to SelectLight grows from 1.65\% to 4.54\%. At scale 2.0, SO-DMPC falls to fourth place on ACQ, ATT, and AWT; GAT-IPPO becomes the strongest baseline, yet SelectLight still lowers ACQ and AWT by 5.57\% and 6.44\%, respectively. SelectLight also retains the lowest ASC at every scale, while its heavy-demand throughput is effectively tied with GAT-IPPO (0.04\% lower). The near-identical throughput shows that SelectLight's high-demand queue and delay reductions are achieved without sacrificing completed trips.

The broader ranking reveals distinct responses to increasing load. GAT-IPPO retains its advantage over IPPO at every scale and becomes the strongest baseline under the heaviest demand, demonstrating that graph-based information exchange remains effective as congestion intensifies. MP exhibits the largest deterioration and remains the weakest method on the delay-related metrics, while MO-DMPC loses competitiveness beyond baseline demand. Actuated control is more stable than MP but continues to trail the learning methods. These patterns highlight the importance of network-aware coordination under high demand.

SelectLight's advantage over SO-DMPC expands sharply with demand. From scale 1.0 to 2.0, the ACQ improvement grows from 1.65\% to 12.35\%, and the AWT improvement grows from 1.76\% to 12.09\%. The combined effect of multi-objective candidate generation and learned plan selection is therefore most valuable as the network becomes congested.

\subsection{Zero-shot transfer}
\label{sec:zero-shot-transfer}

\begin{figure}
\centering
\includegraphics[width=\textwidth]{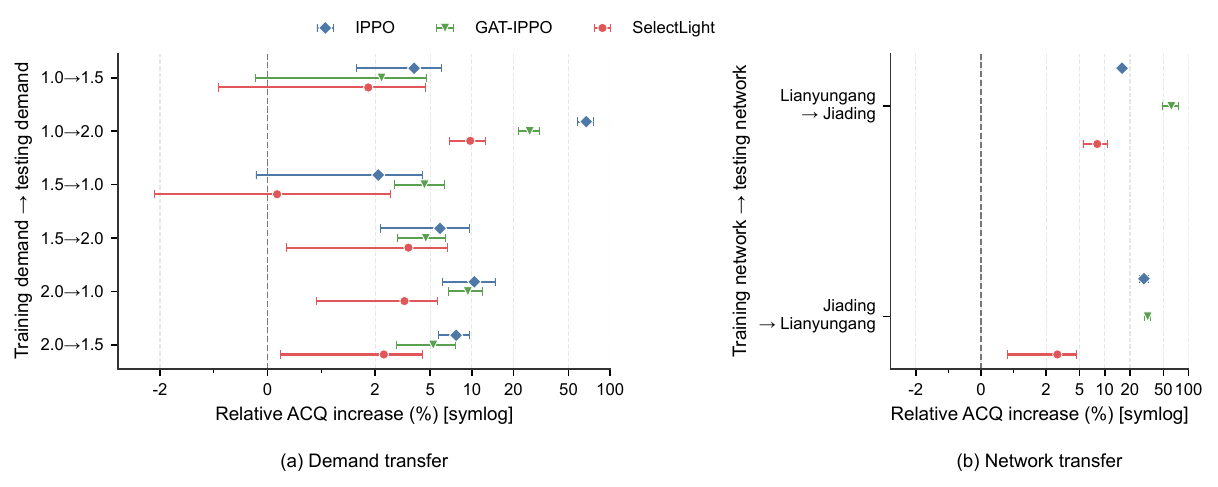}
\caption{Zero-shot transfer without fine-tuning: (a) directed transfer among Lianyungang demand scales and (b) bidirectional transfer between the Lianyungang and Jiading networks. Markers show mean ACQ increases relative to the same method trained in the target domain, with error bars denoting sample standard deviations over $N=8$ seeds. Lower is better. The horizontal axis uses a symmetric logarithmic scale.}
\label{fig:zero-shot-transfer}
\end{figure}

To address RQ2, we test whether a learned controller remains effective when its deployment demand or network differs from training. Demand transfer covers every directed pair among scales 1.0, 1.5, and 2.0 on Lianyungang, while network transfer exchanges the Lianyungang and Jiading policies trained at scale 1.0. SelectLight, IPPO, and GAT-IPPO are evaluated without fine-tuning or test-time adaptation. Transfer loss is the percentage increase in ACQ relative to the same method trained directly in the target domain. Figure~\ref{fig:zero-shot-transfer} summarizes these losses; zero denotes parity with target-domain training, and lower is better.

SelectLight's main demand-transfer advantage is its resistance to a large upward demand shift. It has the lowest transfer loss in all six directions, and under the hardest shift from scale 1.0 to 2.0 its loss is 9.74\%, compared with 26.19\% for GAT-IPPO and 67.11\% for IPPO. Its loss remains between 0.18\% and 3.48\% in the other five directions, demonstrating strong adaptability across both upward and downward demand changes.

SelectLight also degrades less when both topology and demand pattern change. Its transfer loss is 8.14\% from Lianyungang to Jiading, versus 16.07\% for the next-best method, and 2.72\% in the reverse direction, versus 29.35\% for the next-best method. The consistent margin in both directions demonstrates stronger portability across the two urban networks: generating candidate plans online helps SelectLight adapt to a new network without retraining. Complete absolute results are provided in Tables~\ref{tab:app-zero-shot-demand} and~\ref{tab:app-zero-shot-network}.

\subsection{Ablation study}
\label{sec:ablation-study}

To address the component-level aspect of RQ3, the ablation study tests whether performance depends more strongly on predictive policy inputs or candidate-generation objectives (Table~\ref{tab:ablation-study}). Two variants remove either the predicted-arrival profile or the plan-structure inputs listed in Table~\ref{tab:upper-candidate-features}. The latter zeros the stage features and stage mask and removes the phase-change count from the candidate features, while retaining the absolute and relative objective features and the candidate mask. Three further variants disable the delay, queue, or stop objective in SP-MODP; an availability mask distinguishes a disabled objective from a valid zero-valued feature. Every variant is trained from scratch under the common protocol.

\begin{table}
\centering
\caption{Ablation results on the Lianyungang network under demand scale 1.0.}
\label{tab:ablation-study}
\resizebox{\textwidth}{!}{%
\begin{tabular}{lccccc}
\toprule
Variant & ACQ $\downarrow$ & ATT $\downarrow$ & AWT $\downarrow$ & ASC $\downarrow$ & THP $\uparrow$ \\
\midrule
Full & {\bfseries\boldmath $9720.03 \pm 77.03$} & {\bfseries\boldmath $283.38 \pm 1.47$} & {\bfseries\boldmath $42.10 \pm 0.35$} & {\bfseries\boldmath $2.494 \pm 0.017$} & {\bfseries\boldmath $5851.62 \pm 14.28$} \\
w/o predicted arrivals & \underline{$9809.19 \pm 107.36$} & \underline{$283.60 \pm 1.98$} & \underline{$42.45 \pm 0.41$} & \underline{$2.499 \pm 0.031$} & \underline{$5849.38 \pm 14.84$} \\
w/o plan structure & $9923.02 \pm 221.73$ & $284.43 \pm 2.08$ & $42.86 \pm 0.82$ & $2.513 \pm 0.033$ & $5840.00 \pm 17.25$ \\
w/o delay objective & $10970.02 \pm 258.47$ & $288.18 \pm 3.35$ & $47.50 \pm 1.22$ & $2.503 \pm 0.036$ & $5829.62 \pm 9.38$ \\
w/o queue objective & $9974.73 \pm 95.45$ & $284.83 \pm 1.26$ & $43.28 \pm 0.36$ & $2.509 \pm 0.023$ & $5848.12 \pm 11.05$ \\
w/o stops objective & $9895.86 \pm 77.40$ & $284.63 \pm 1.18$ & $42.90 \pm 0.34$ & $2.530 \pm 0.029$ & $5846.38 \pm 10.20$ \\
\bottomrule
\end{tabular}}
\end{table}

The delay objective is the dominant contributor among the tested components: removing it increases ACQ and AWT by 12.86\% and 12.83\%, respectively. Every other removal changes ACQ by less than 3\%, with the queue objective and plan structure having the next-largest effects. Explicit delay optimization therefore establishes the main performance signal, while the remaining objectives and predictive features refine the selected plans.

\subsection{Operational interpretation}
\label{sec:operational-interpretation}

We address the operational aspect of RQ3 by examining corridor progression and candidate-selection behavior.

\subsubsection{Corridor progression}
\label{sec:corridor-progression}

\begin{figure}
\centering
\includegraphics[width=\textwidth]{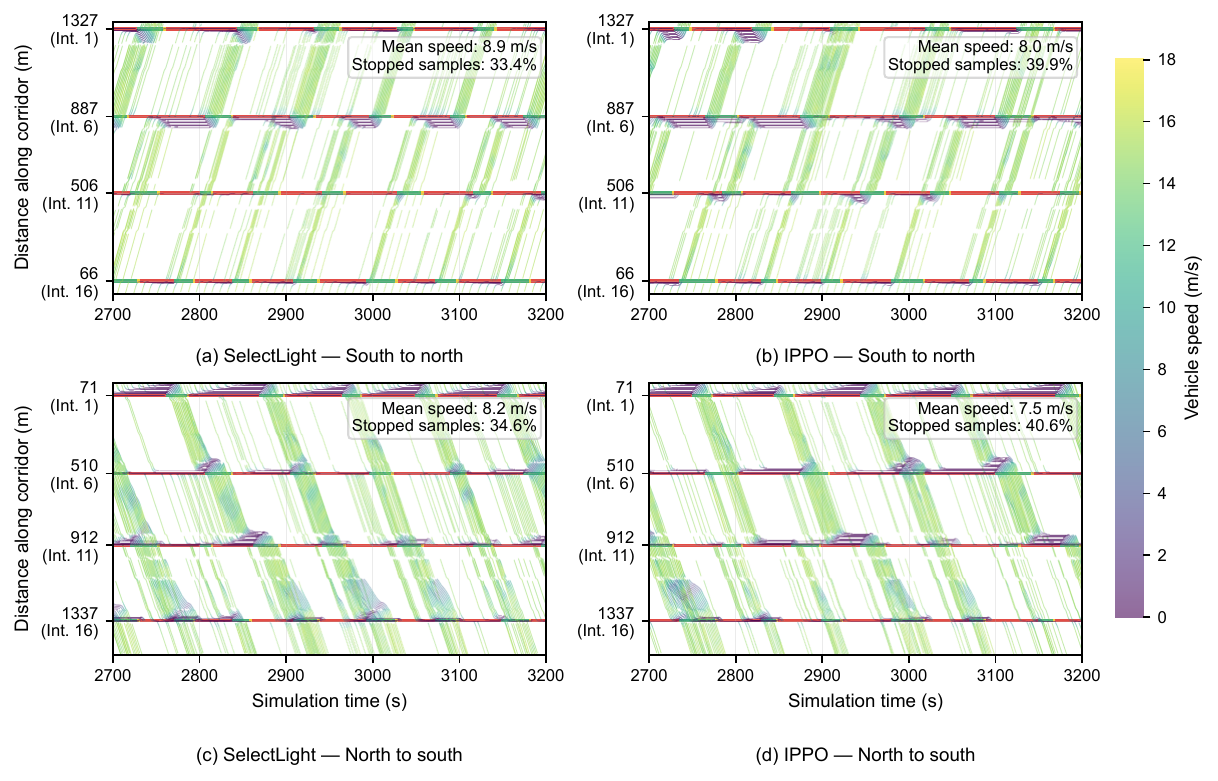}
\caption{Vehicle trajectories under SelectLight and IPPO on a representative Lianyungang corridor during 2700--3200~s at demand scale 2.0. Panels (a) and (b) follow Int.~16--11--6--1 from south to north, while panels (c) and (d) follow Int.~1--6--11--16 from north to south. Mean speeds and shares of stopped trajectory samples (speed below 1~m/s) refer to the displayed corridor and time window.}
\label{fig:corridor-progression}
\end{figure}

To examine whether the network-level delay reduction corresponds to smoother traffic progression, Figure~\ref{fig:corridor-progression} compares SelectLight and IPPO on the same north--south corridor at demand scale 2.0. IPPO is the strongest baseline without policy-level inter-agent feature exchange at this demand level, ranking behind only GAT-IPPO on ACQ, ATT, and AWT. It shares SelectLight's PPO setting and topology-aware local encoder but directly selects phase switches from current detector observations, providing a focused reference for the progression gains of predictive candidate-plan control.

The trajectory window spans 2700--3200~s. Horizontal red and green segments show the major-direction signal state, trajectory color indicates instantaneous speed, and near-horizontal trajectories indicate stopped or slow-moving vehicles. South-to-north traffic passes Int.~16, Int.~11, Int.~6, and Int.~1 in sequence; north-to-south traffic follows the reverse order.

The time--space diagrams reveal different progression patterns in the two directions. From south to north, SelectLight maintains largely uninterrupted movement along the corridor, with pronounced queueing concentrated upstream of Int.~6. Under IPPO, smaller stopped clusters also emerge at Int.~11 and Int.~1, spreading interruptions across multiple intersections. From north to south, SelectLight forms a recognizable green-wave pattern in approximately half of the visible cycles: platoons released from Int.~1 continue through Int.~6, Int.~11, and Int.~16 with limited stopping. The IPPO trajectories are repeatedly fragmented at downstream intersections and show no comparably regular progression pattern.

These progression patterns are reflected in the corridor summaries. From south to north, SelectLight raises mean speed from 8.0 to 8.9~m/s and reduces the stopped-sample share from 39.9\% to 33.4\%. From north to south, mean speed rises from 7.5 to 8.2~m/s and the stopped-sample share falls from 40.6\% to 34.6\%. Together, these results demonstrate SelectLight's stronger corridor-level coordination, with fewer stop-and-go interruptions and more continuous progression across successive intersections.

\subsubsection{Candidate-selection behavior}
\label{sec:operational-candidate-selection}

To examine how the learned selector uses the candidate set, we compare each SelectLight decision with two rules applied to the same candidates and traffic state. The ideal-point rule first scales delay, queue, and stops independently to $[0,1]$ within the current set, and then orders candidates by their Euclidean distance to the ideal point $(0,0,0)$. Its first-ranked candidate is the fixed choice made by MO-DMPC. The minimum-delay rule instead selects the candidate with the smallest predicted queueing delay. Here, this rule operates within SelectLight's multi-objective candidate set; SO-DMPC in Table~\ref{tab:overall-performance} generates delay-only candidates throughout closed-loop control.

\begin{figure}
\centering
\includegraphics[width=\textwidth]{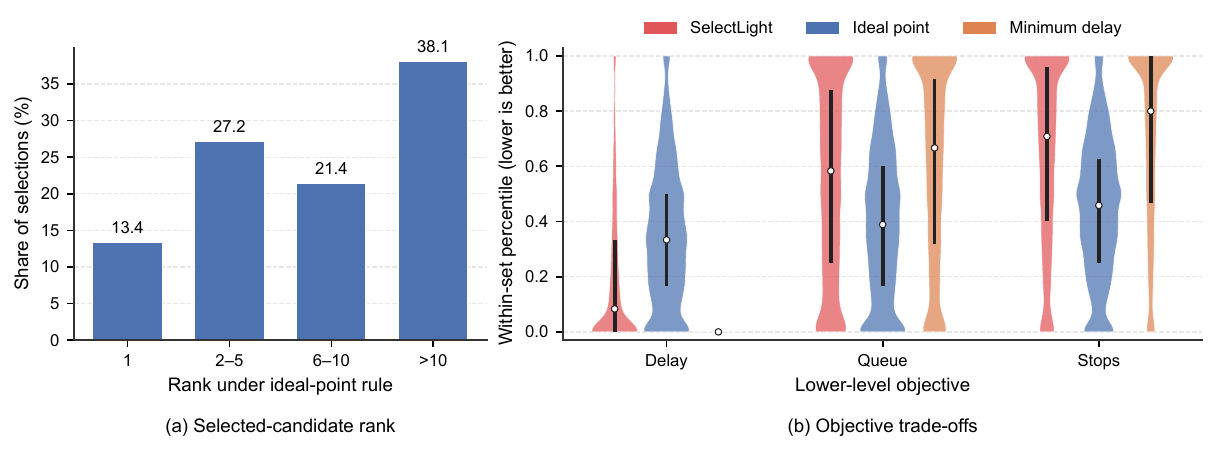}
\caption{Candidate-selection behavior on the Lianyungang network at demand scale 2.0: (a) distribution of the ideal-point rank of each SelectLight choice, where rank~1 is the candidate selected by the fixed ideal-point rule; and (b) distributions of within-set objective percentiles for SelectLight and the two comparison rules. Lower percentiles are better. White dots denote medians and black segments denote interquartile ranges. Statistics contain 161,280 candidate sets from $N=8$ evaluation seeds.}
\label{fig:candidate-selection-behavior}
\end{figure}

Figure~\ref{fig:candidate-selection-behavior}(a) measures how closely SelectLight follows the ideal-point ordering. The \emph{ideal-point rank} is the one-based position of the selected candidate after sorting all candidates by increasing distance to $(0,0,0)$: rank~1 means that SelectLight chooses exactly the ideal-point candidate, while a higher rank indicates a larger departure from that fixed rule. It measures agreement with the heuristic, not the overall quality of a candidate. SelectLight chooses rank~1 in only 13.38\% of the candidate sets, whereas 59.43\% of its choices have rank~6 or higher. Thus, its decisions frequently differ substantially from the fixed ideal-point choice.

Figure~\ref{fig:candidate-selection-behavior}(b) shows which objectives these departures favor. For each objective separately, candidates are ordered from best to worst within the current set and their positions are scaled from 0 to 1. A percentile of 0 therefore denotes the best available candidate for that objective, and 1 the worst. Each violin shows the distribution across candidate sets; the white dot marks the median and the black segment the interquartile range.

As shown in the figure, SelectLight operates between the two fixed rules, with a clear preference for reducing delay. Compared with the ideal-point rule, SelectLight lowers the median delay percentile from 0.333 to 0.083, while accepting higher queue and stop percentiles. Minimum-delay selection always attains a delay percentile of 0 by construction, but its queue and stop percentiles are more concentrated toward poorer values. SelectLight gives up this single-objective optimum to improve the median queue percentile from 0.667 to 0.583 and the stop percentile from 0.800 to 0.708. The learned selector therefore adapts the balance among all three objectives instead of reproducing either the fixed ideal-point rule or the delay-only rule.

\subsection{Computational efficiency}
\label{sec:computational-efficiency}

To address RQ4, we assess whether online candidate generation fits within the control cycle. We measure SP-MODP latency using one SUMO worker, one SP-MODP solver thread, and CPU execution over a complete 3600~s Lianyungang episode at demand scale 1.0. Table~\ref{tab:solver-runtime} reports per-intersection solution times for four prediction horizons.

\begin{table}
\centering
\caption{Serial SP-MODP solution time per intersection across prediction horizons, based on 20,160 solves per horizon. The dagger marks the default setting.}
\label{tab:solver-runtime}
\begin{tabular}{cccc}
\toprule
$H$ (s) & Median (ms) & p95 (ms) & p99 (ms) \\
\midrule
60 & 0.139 & 0.337 & 0.521 \\
90 & 0.342 & 1.052 & 1.927 \\
120\textsuperscript{$\dagger$} & 0.749 & 2.993 & 5.408 \\
150 & 1.575 & 7.770 & 14.158 \\
\bottomrule
\end{tabular}
\end{table}

SP-MODP computation is comfortably within the online control budget at the default setting. For $H=120$~s, the p99 latency is 5.408~ms per intersection, compared with the 5~s macro decision interval. Extending the horizon to 150~s raises the median latency from 0.749 to 1.575~ms and the p99 to 14.158~ms, while the sensitivity study in \ref{app:parameter-sensitivity} finds small, metric-dependent performance changes. The default $H=120$~s therefore combines competitive control performance with lower computational cost.

\section{Conclusion}
\label{sec:conclusion}

This paper introduced post-optimization selection as a new MPC--RL integration pattern and instantiated it in SelectLight for network traffic signal control. Rather than using reinforcement learning to configure an optimizer or correct its output, SelectLight uses distributed model predictive control to construct bounded sets of feasible signal plans and gives a learned selector the final choice among them. SP-MODP preserves trade-offs among three predicted objectives---total queueing delay, peak queue accumulation, and total number of stops---while the topology-aware attention policy accommodates heterogeneous intersections and variable-size candidate sets. This division of responsibility retains online predictive planning and preserves signal timing feasibility by construction, yet allows plan selection to be learned from realized closed-loop outcomes.

Experiments on two 28-intersection urban networks support this design. SelectLight achieved the lowest queueing delay, travel time, and waiting time on both networks; relative to the strongest DMPC baseline, it reduced queueing delay and waiting time by 1.68--1.95\% under baseline demand. Its advantage increased as congestion intensified: at twice the baseline demand, it reduced queueing delay and waiting time by 5.57\% and 6.44\%, respectively, relative to the strongest competing method, with essentially unchanged throughput. SelectLight also yielded the lowest transfer loss in every tested demand-transfer direction and degraded less than direct RL controllers when transferred between networks without fine-tuning. Ablation and selection-behavior analyses further showed that predicted delay provides the dominant optimization signal, while learned selection uses the remaining objectives and plan structure to depart systematically from fixed ideal-point and minimum-delay rules. At the default 120~s horizon, the per-intersection p99 SP-MODP latency was 5.408~ms, well below the 5~s control interval. Together, these results show that learning over optimizer-generated action sets can combine the adaptability of RL with the feasibility and online replanning of MPC without requiring RL to synthesize signal plans directly.

The present study is limited to microscopic simulation on two networks with fixed path-flow demands and assumes a calibrated, stationary AVI-based propagation model. Possible extensions include (i) accounting for prediction and sensing uncertainty in candidate generation and selection, (ii) evaluating time-varying demand, incidents, missing observations, and communication failures, (iii) scaling the framework to larger networks, and (iv) conducting end-to-end real-time and field evaluations.

\section*{Declaration of generative AI and AI-assisted technologies in the writing process}
During the preparation of this work, the authors used ChatGPT in order to improve the language. After using this tool/service, the authors reviewed and edited the content as needed and take full responsibility for the content of the publication.

\section*{CRediT authorship contribution statement}
\textbf{Lyuzhou Luo:} Conceptualization, Methodology, Software, Validation, Formal analysis, Investigation, Data curation, Writing---original draft, Visualization; \textbf{Chaopeng Tan:} Conceptualization, Methodology, Validation, Writing---review \& editing, Supervision; \textbf{Zhengyong Gao:} Validation, Investigation, Data curation; \textbf{Hong Zhu:} Writing---review \& editing, Supervision; \textbf{Andrea D'Ariano:} Writing---review \& editing, Supervision; \textbf{Keshuang Tang:} Conceptualization, Writing---review \& editing, Supervision, Project administration, Funding acquisition.

\section*{Data availability}
The data that support the findings of this study are available from the authors on reasonable request.

\section*{Acknowledgment}
This work was supported by the National Key Research \& Development Program of China under Grant 2023YFB4301900; the Guangxi Key Research and Development Program under Grant AB25069483; the National Natural Science Foundation of China under Grants 52572360, 52372319, and 52302414; the Key International (Regional) Joint Research Program of the National Natural Science Foundation of China under Grant W2511048; and the 2024 Shanghai ``Science and Technology Innovation Action Plan'' International Scientific and Technological Cooperation Program under Grant 24510714400.

\appendix
\renewcommand{\tablebodyfont}{\footnotesize}

\section{Implementation and training details}
\label{app:proposed-method}

\subsection{Learning-method implementations}
\label{app:learning-method-implementation}

\subsubsection{SelectLight}
\label{app:selectlight-implementation}

The implementation follows the observation, encoding, and candidate-scoring definitions in Section~\ref{sec:upper-level-rl-policy}. The actor and critic use separate parameter sets, while each network shares its parameters across intersections. Variable-size lane, phase, stage, and candidate sets are padded only for batching. Their validity masks are retained throughout encoding, pooling, action scoring, and value aggregation so that padded entries do not affect the policy or value estimate.

\subsubsection{Direct RL baselines}
\label{app:direct-rl-baselines}

IPPO and GAT-IPPO use detector-based current observations and directly select phase-switching actions. They do not receive predictive arrival profiles, candidate plan features, or SP-MODP outputs. Their observations contain current lane queues and densities, signal timing, and the static lane--phase service relations required by the topology-aware local encoder. Actor and critic encoders are separate, and each network shares its parameters across intersections.

Both baselines use a binary action: retain the current phase or switch to the next phase in the native cyclic sequence. Before the minimum green is reached, only retention is valid; at the maximum green, only switching is valid. At intermediate times, both actions are available. A switch executes the 3~s intergreen followed by 2~s of the next green within the 5~s macro decision interval.

Table~\ref{tab:learning-method-structures} summarizes how the two direct RL baselines differ structurally from SelectLight. IPPO maps the local intersection embedding through separate MLP policy and value heads. GAT-IPPO first applies the same local encoder and then exchanges intersection embeddings along the direct network adjacency graph through separate actor and critic graph-attention blocks, each configured with two layers and four attention heads. The three methods share the same reward, PPO protocol, and signal constraints. They differ in their observations, network blocks, and action spaces.

\begin{table}
\centering
\caption{Structural differences among the learning methods.}
\label{tab:learning-method-structures}
\begin{tabular}{lccc}
\toprule
Component & SelectLight & IPPO & GAT-IPPO \\
\midrule
Topology-aware local encoder & Yes & Yes & Yes \\
Current detector state & Yes & Yes & Yes \\
Predictive arrival profile & Yes & No & No \\
Candidate plan features & Yes & No & No \\
Inter-agent graph attention & No & No & Yes \\
Action & Candidate plan index & Retain/switch & Retain/switch \\
\bottomrule
\end{tabular}
\end{table}

\subsection{Training protocol}
\label{app:training-details}

SelectLight, IPPO, and GAT-IPPO use the common PPO protocol in Table~\ref{tab:ppo-settings}. Each method is trained independently, and the actor and critic use separate AdamW optimizers.

\begin{table}
\centering
\caption{PPO training configuration shared by the learning methods.}
\label{tab:ppo-settings}
\begin{tabular}{ll}
\toprule
Hyperparameter & Value \\
\midrule
Optimizer & AdamW \\
Actor / critic learning rate & $3\times10^{-4}$ / $5\times10^{-4}$ \\
Discount factor $\gamma$ & 0.99 \\
GAE parameter $\lambda$ & 0.95 \\
PPO clipping parameter & 0.2 \\
Entropy coefficient & 0.01 \\
Approximate KL early-stopping threshold & 0.01 \\
Actor / critic maximum gradient norm & 1.0 / 1.0 \\
Critic objective & Smooth L1 with PopArt-normalized targets \\
PPO epochs per rollout & 10 \\
Parallel environments & 32 \\
Frames per rollout batch & 5,760 \\
Minibatches per rollout & 4 \\
Total training frames & 2,004,480 \\
Learning-rate schedule & 2\% warmup, then linear decay to 10\% of peak \\
\bottomrule
\end{tabular}
\end{table}

Rollout workers are deterministically staggered across episode phases to avoid collecting synchronized fragments from the same part of an episode. Advantages are standardized per intersection. Both networks use the same training protocol. Their simulation and observation inputs are network specific.

\FloatBarrier
\section{Additional experimental results}
\label{app:additional-experimental-results}

\subsection{Training curves}
\label{app:training-curves}

\begin{figure}
\centering
\includegraphics[width=0.8\textwidth]{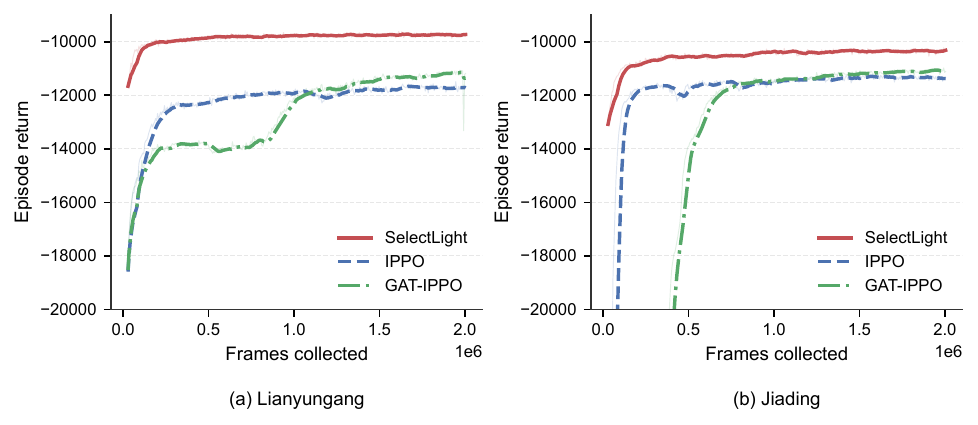}
\caption{Training curves of SelectLight, IPPO, and GAT-IPPO on the Lianyungang and Jiading networks at the baseline demand. Thick lines are ten-point moving averages of the environment-averaged episode returns.}
\label{fig:training-curves}
\end{figure}

Figure~\ref{fig:training-curves} shows the final training run for each learning method and network at the baseline demand. Episode return is the accumulated negative realized local queueing delay, so a higher return indicates less accumulated delay. Thin lines show the recorded returns after averaging across parallel environments; thick lines show ten-point moving averages. Each curve represents one training run and characterizes within-run optimization behavior.

SelectLight reaches a higher return plateau earlier than the direct-action baselines on both networks. The consistent pattern indicates that the candidate-plan action space provides a more structured learning problem than direct phase switching.

\FloatBarrier
\subsection{Complete results across demand levels}
\label{app:demand-sensitivity-results}

Table~\ref{tab:app-demand-robustness} complements Fig.~\ref{fig:demand-robustness} by reporting all five metrics for every controller and demand scale. The complete results make the stop-count and throughput trade-offs visible alongside the three delay-related metrics emphasized in the main text.

\begin{table}
\centering
\caption{Complete performance results across demand levels on the Lianyungang network. All methods are evaluated at the indicated demand scale; each learning method is trained at the same scale.}
\label{tab:app-demand-robustness}
\resizebox{\textwidth}{!}{%
\begin{tabular}{llccccc}
\toprule
Demand scale & Method & ACQ $\downarrow$ & ATT $\downarrow$ & AWT $\downarrow$ & ASC $\downarrow$ & THP $\uparrow$ \\
\midrule
\multirow{7}{*}{1.0} & Actuated & $12532.57 \pm 201.77$ & $300.22 \pm 1.41$ & $54.48 \pm 0.86$ & $3.202 \pm 0.042$ & $5827.25 \pm 18.43$ \\
 & MP & $16192.09 \pm 227.21$ & $314.02 \pm 2.38$ & $70.37 \pm 1.00$ & $3.072 \pm 0.025$ & $5786.50 \pm 12.73$ \\
 & MO-DMPC & $11631.25 \pm 295.06$ & $292.84 \pm 1.93$ & $50.66 \pm 1.35$ & \underline{$2.545 \pm 0.014$} & $5837.38 \pm 15.49$ \\
 & SO-DMPC & \underline{$9886.19 \pm 159.47$} & \underline{$284.61 \pm 1.67$} & \underline{$42.86 \pm 0.69$} & $2.570 \pm 0.030$ & \underline{$5849.12 \pm 9.86$} \\
 & IPPO & $11683.03 \pm 248.63$ & $293.31 \pm 2.26$ & $50.74 \pm 1.06$ & $2.734 \pm 0.038$ & $5830.62 \pm 12.34$ \\
 & GAT-IPPO & $11114.40 \pm 179.16$ & $290.25 \pm 2.32$ & $48.15 \pm 0.79$ & $2.695 \pm 0.047$ & $5840.50 \pm 11.51$ \\
 & SelectLight & {\bfseries\boldmath $9720.03 \pm 77.03$} & {\bfseries\boldmath $283.38 \pm 1.47$} & {\bfseries\boldmath $42.10 \pm 0.35$} & {\bfseries\boldmath $2.494 \pm 0.017$} & {\bfseries\boldmath $5851.62 \pm 14.28$} \\
\midrule
\multirow{7}{*}{1.5} & Actuated & $21643.85 \pm 390.11$ & $311.36 \pm 1.75$ & $62.26 \pm 1.04$ & $3.429 \pm 0.034$ & $8691.00 \pm 20.59$ \\
 & MP & $28627.41 \pm 509.26$ & $330.93 \pm 3.52$ & $82.47 \pm 1.62$ & $3.486 \pm 0.055$ & $8621.50 \pm 17.00$ \\
 & MO-DMPC & $20802.12 \pm 363.58$ & $303.75 \pm 2.27$ & $60.24 \pm 1.07$ & \underline{$2.790 \pm 0.032$} & $8692.12 \pm 19.77$ \\
 & SO-DMPC & \underline{$17404.12 \pm 154.06$} & \underline{$295.55 \pm 1.52$} & \underline{$50.12 \pm 0.45$} & $2.818 \pm 0.017$ & \underline{$8727.75 \pm 18.31$} \\
 & IPPO & $19465.91 \pm 326.34$ & $301.70 \pm 2.00$ & $56.06 \pm 1.01$ & $2.996 \pm 0.021$ & $8713.25 \pm 23.43$ \\
 & GAT-IPPO & $18776.99 \pm 292.83$ & $299.76 \pm 2.83$ & $53.98 \pm 0.96$ & $2.922 \pm 0.055$ & $8723.25 \pm 17.90$ \\
 & SelectLight & {\bfseries\boldmath $16613.33 \pm 415.54$} & {\bfseries\boldmath $291.93 \pm 3.05$} & {\bfseries\boldmath $47.72 \pm 1.23$} & {\bfseries\boldmath $2.718 \pm 0.045$} & {\bfseries\boldmath $8741.25 \pm 19.01$} \\
\midrule
\multirow{7}{*}{2.0} & Actuated & $38586.39 \pm 1130.24$ & $336.54 \pm 3.01$ & $83.41 \pm 2.17$ & $3.732 \pm 0.056$ & $11451.62 \pm 40.61$ \\
 & MP & $51922.12 \pm 538.29$ & $367.55 \pm 1.71$ & $109.21 \pm 1.24$ & $4.477 \pm 0.043$ & $11287.00 \pm 33.14$ \\
 & MO-DMPC & $39952.53 \pm 431.71$ & $335.55 \pm 1.72$ & $86.11 \pm 1.14$ & \underline{$3.293 \pm 0.029$} & $11405.38 \pm 48.47$ \\
 & SO-DMPC & $34579.30 \pm 1194.36$ & $323.93 \pm 2.58$ & $73.38 \pm 1.88$ & $3.299 \pm 0.051$ & $11419.75 \pm 37.90$ \\
 & IPPO & $33118.38 \pm 695.59$ & $321.28 \pm 2.44$ & $71.29 \pm 1.48$ & $3.333 \pm 0.048$ & $11524.00 \pm 36.08$ \\
 & GAT-IPPO & \underline{$32073.92 \pm 290.77$} & \underline{$319.29 \pm 1.40$} & \underline{$68.93 \pm 0.62$} & $3.329 \pm 0.026$ & {\bfseries\boldmath $11535.75 \pm 20.89$} \\
 & SelectLight & {\bfseries\boldmath $30286.20 \pm 1053.24$} & {\bfseries\boldmath $313.77 \pm 2.85$} & {\bfseries\boldmath $64.49 \pm 2.22$} & {\bfseries\boldmath $3.172 \pm 0.048$} & \underline{$11531.38 \pm 17.84$} \\
\bottomrule
\end{tabular}}
\end{table}

\FloatBarrier
\subsection{Complete zero-shot transfer results}
\label{app:zero-shot-results}

The complete transfer tables show whether the ACQ losses summarized in Fig.~\ref{fig:zero-shot-transfer} coincide with changes in the other traffic metrics. Tables~\ref{tab:app-zero-shot-demand} and~\ref{tab:app-zero-shot-network} report the absolute five-metric results for policies evaluated without fine-tuning or test-time adaptation. The target-trained references appear in Table~\ref{tab:app-demand-robustness}, so the demand-transfer table contains the 18 off-diagonal train--test combinations.

\begin{table}
\centering
\caption{Complete zero-shot demand-transfer results on the Lianyungang network without fine-tuning or test-time adaptation.}
\label{tab:app-zero-shot-demand}
\resizebox{\textwidth}{!}{%
\begin{tabular}{lllccccc}
\toprule
Test scale & Train scale & Method & ACQ $\downarrow$ & ATT $\downarrow$ & AWT $\downarrow$ & ASC $\downarrow$ & THP $\uparrow$ \\
\midrule
\multirow{6}{*}{1.0} & 1.5 & SelectLight & {\bfseries\boldmath $9736.29 \pm 191.48$} & \underline{$283.70 \pm 2.26$} & {\bfseries\boldmath $42.24 \pm 0.83$} & {\bfseries\boldmath $2.484 \pm 0.043$} & {\bfseries\boldmath $5850.12 \pm 7.40$} \\
 & 1.5 & IPPO & $11924.95 \pm 161.00$ & $294.71 \pm 1.25$ & $51.75 \pm 0.64$ & $2.766 \pm 0.034$ & $5831.38 \pm 8.16$ \\
 & 1.5 & GAT-IPPO & $11618.03 \pm 174.03$ & $293.85 \pm 2.62$ & $50.33 \pm 0.81$ & $2.748 \pm 0.029$ & $5831.88 \pm 12.94$ \\
 & 2.0 & SelectLight & \underline{$10036.01 \pm 201.52$} & {\bfseries\boldmath $283.64 \pm 2.85$} & \underline{$43.51 \pm 0.91$} & \underline{$2.514 \pm 0.048$} & \underline{$5849.00 \pm 8.35$} \\
 & 2.0 & IPPO & $12892.65 \pm 269.53$ & $299.71 \pm 2.34$ & $55.95 \pm 1.12$ & $2.832 \pm 0.036$ & $5813.12 \pm 18.10$ \\
 & 2.0 & GAT-IPPO & $12151.53 \pm 174.72$ & $295.75 \pm 1.75$ & $52.71 \pm 0.72$ & $2.784 \pm 0.031$ & $5829.88 \pm 18.22$ \\
\midrule
\multirow{6}{*}{1.5} & 1.0 & SelectLight & {\bfseries\boldmath $16914.58 \pm 108.49$} & {\bfseries\boldmath $292.39 \pm 1.31$} & {\bfseries\boldmath $48.61 \pm 0.42$} & \underline{$2.739 \pm 0.031$} & {\bfseries\boldmath $8742.38 \pm 16.64$} \\
 & 1.0 & IPPO & $20206.64 \pm 323.61$ & $304.33 \pm 1.93$ & $58.16 \pm 0.98$ & $3.039 \pm 0.035$ & $8701.38 \pm 24.33$ \\
 & 1.0 & GAT-IPPO & $19190.57 \pm 329.33$ & $301.68 \pm 3.53$ & $55.35 \pm 1.05$ & $2.960 \pm 0.052$ & $8726.75 \pm 13.84$ \\
 & 2.0 & SelectLight & \underline{$16990.62 \pm 170.51$} & \underline{$292.39 \pm 2.02$} & \underline{$48.92 \pm 0.56$} & {\bfseries\boldmath $2.723 \pm 0.024$} & \underline{$8735.25 \pm 17.04$} \\
 & 2.0 & IPPO & $20959.12 \pm 252.27$ & $305.57 \pm 1.44$ & $60.46 \pm 0.68$ & $3.003 \pm 0.030$ & $8698.38 \pm 13.59$ \\
 & 2.0 & GAT-IPPO & $19759.75 \pm 307.57$ & $302.53 \pm 1.97$ & $56.77 \pm 0.89$ & $2.972 \pm 0.036$ & $8688.62 \pm 17.10$ \\
\midrule
\multirow{6}{*}{2.0} & 1.0 & SelectLight & \underline{$33215.74 \pm 743.20$} & \underline{$321.36 \pm 2.99$} & \underline{$70.35 \pm 1.65$} & \underline{$3.322 \pm 0.065$} & $11478.12 \pm 19.95$ \\
 & 1.0 & IPPO & $55344.48 \pm 3155.33$ & $369.93 \pm 6.75$ & $107.36 \pm 4.99$ & $4.886 \pm 0.238$ & $10965.50 \pm 98.52$ \\
 & 1.0 & GAT-IPPO & $40467.82 \pm 1205.02$ & $338.83 \pm 4.08$ & $84.87 \pm 2.82$ & $3.736 \pm 0.104$ & $11326.75 \pm 36.33$ \\
 & 1.5 & SelectLight & {\bfseries\boldmath $31318.02 \pm 664.27$} & {\bfseries\boldmath $316.25 \pm 2.20$} & {\bfseries\boldmath $66.25 \pm 1.44$} & {\bfseries\boldmath $3.258 \pm 0.033$} & \underline{$11503.75 \pm 18.08$} \\
 & 1.5 & IPPO & $35052.18 \pm 866.65$ & $327.53 \pm 4.17$ & $75.37 \pm 1.82$ & $3.543 \pm 0.075$ & $11496.75 \pm 24.38$ \\
 & 1.5 & GAT-IPPO & $33563.00 \pm 560.13$ & $323.10 \pm 2.94$ & $71.89 \pm 1.41$ & $3.443 \pm 0.059$ & {\bfseries\boldmath $11513.25 \pm 31.79$} \\
\bottomrule
\end{tabular}}
\end{table}

\begin{table}
\centering
\caption{Complete zero-shot network-transfer results without fine-tuning or test-time adaptation.}
\label{tab:app-zero-shot-network}
\resizebox{\textwidth}{!}{%
\begin{tabular}{lllccccc}
\toprule
Train network & Test network & Method & ACQ $\downarrow$ & ATT $\downarrow$ & AWT $\downarrow$ & ASC $\downarrow$ & THP $\uparrow$ \\
\midrule
\multirow{3}{*}{Lianyungang} & \multirow{3}{*}{Jiading} & SelectLight & {\bfseries\boldmath $11088.00 \pm 192.22$} & {\bfseries\boldmath $328.88 \pm 1.48$} & {\bfseries\boldmath $62.67 \pm 1.10$} & {\bfseries\boldmath $3.168 \pm 0.033$} & {\bfseries\boldmath $4398.75 \pm 4.65$} \\
 &  & IPPO & \underline{$13181.86 \pm 182.48$} & \underline{$341.52 \pm 1.46$} & \underline{$74.59 \pm 1.06$} & \underline{$3.334 \pm 0.035$} & \underline{$4376.12 \pm 6.01$} \\
 &  & GAT-IPPO & $17785.89 \pm 1390.15$ & $368.49 \pm 6.39$ & $93.48 \pm 4.71$ & $4.396 \pm 0.210$ & $4274.38 \pm 34.76$ \\
\midrule
\multirow{3}{*}{Jiading} & \multirow{3}{*}{Lianyungang} & SelectLight & {\bfseries\boldmath $9983.92 \pm 144.67$} & {\bfseries\boldmath $282.83 \pm 2.49$} & {\bfseries\boldmath $43.21 \pm 0.67$} & {\bfseries\boldmath $2.509 \pm 0.032$} & {\bfseries\boldmath $5849.00 \pm 10.89$} \\
 &  & IPPO & $15106.75 \pm 269.62$ & $310.06 \pm 3.42$ & $65.95 \pm 1.24$ & \underline{$2.925 \pm 0.039$} & \underline{$5804.25 \pm 19.60$} \\
 &  & GAT-IPPO & \underline{$14723.84 \pm 96.71$} & \underline{$308.25 \pm 1.63$} & \underline{$64.23 \pm 0.41$} & $2.973 \pm 0.023$ & $5803.00 \pm 11.48$ \\
\bottomrule
\end{tabular}}
\end{table}

\FloatBarrier
\subsection{Parameter sensitivity}
\label{app:parameter-sensitivity}
\label{app:parameter-sensitivity-results}

The sensitivity analysis tests whether SelectLight depends on narrowly tuned planning settings. Each prediction-horizon and candidate-set configuration uses a separately trained policy. When varying $H$, we fix $K=25$; when varying $K$, we fix $H=120$~s. Table~\ref{tab:app-parameter-sensitivity} reports all five performance metrics for the tested settings.

\begin{table}
\centering
\caption{Complete parameter-sensitivity results on the Lianyungang network under demand scale 1.0. Daggers mark the default settings.}
\label{tab:app-parameter-sensitivity}
\resizebox{\textwidth}{!}{%
\begin{tabular}{llccccc}
\toprule
Parameter & Setting & ACQ $\downarrow$ & ATT $\downarrow$ & AWT $\downarrow$ & ASC $\downarrow$ & THP $\uparrow$ \\
\midrule
\multirow{4}{*}{Prediction horizon $H$ (s)} & 60 & $9815.61 \pm 170.58$ & \underline{$282.65 \pm 2.35$} & $42.54 \pm 0.71$ & {\bfseries\boldmath $2.455 \pm 0.036$} & $5851.25 \pm 10.39$ \\
 & 90 & $9763.38 \pm 89.50$ & $283.67 \pm 1.49$ & $42.28 \pm 0.42$ & $2.484 \pm 0.027$ & $5849.25 \pm 11.67$ \\
 & 120\textsuperscript{$\dagger$} & \underline{$9720.03 \pm 77.03$} & $283.38 \pm 1.47$ & \underline{$42.10 \pm 0.35$} & $2.494 \pm 0.017$ & \underline{$5851.62 \pm 14.28$} \\
 & 150 & {\bfseries\boldmath $9712.05 \pm 154.93$} & {\bfseries\boldmath $282.57 \pm 2.09$} & {\bfseries\boldmath $42.04 \pm 0.67$} & \underline{$2.475 \pm 0.031$} & {\bfseries\boldmath $5853.25 \pm 17.77$} \\
\midrule
\multirow{3}{*}{Maximum candidate set size $K$} & 10 & $9759.92 \pm 107.46$ & $283.51 \pm 1.75$ & $42.26 \pm 0.45$ & \underline{$2.492 \pm 0.023$} & {\bfseries\boldmath $5854.00 \pm 8.42$} \\
 & 25\textsuperscript{$\dagger$} & {\bfseries\boldmath $9720.03 \pm 77.03$} & \underline{$283.38 \pm 1.47$} & \underline{$42.10 \pm 0.35$} & $2.494 \pm 0.017$ & \underline{$5851.62 \pm 14.28$} \\
 & 50 & \underline{$9725.98 \pm 83.73$} & {\bfseries\boldmath $282.55 \pm 1.41$} & {\bfseries\boldmath $42.09 \pm 0.39$} & {\bfseries\boldmath $2.476 \pm 0.026$} & $5845.38 \pm 18.48$ \\
\bottomrule
\end{tabular}}
\end{table}

Performance remains stable around the default settings, and the gains from larger settings are small and metric-dependent. Extending the horizon to 150~s lowers ACQ by 0.08\% while approximately doubling median SP-MODP latency, and increasing $K$ from 25 to 50 produces small gains on ATT, AWT, and ASC while doubling the maximum number of candidate slots processed by the selector. The default $H=120$~s and $K=25$ therefore retain competitive performance with lower computational cost and a smaller input size.

\FloatBarrier

\clearpage
\bibliographystyle{elsarticle-harv}
\bibliography{refs}

\end{document}